\documentclass[letterpaper,twocolumn,superscriptaddress,floatfix]{revtex4}
\usepackage[latin1]{inputenc}
\usepackage{bm}
\usepackage{multirow,amssymb,amsbsy,amsmath}
\usepackage{graphicx}
\usepackage{verbatim}
\makeatletter
\usepackage{pifont}
\makeatother

\begin{document}

\title{Experimental fault-tolerant universal quantum gates with solid-state spins under ambient conditions}

\author{Xing Rong}
\thanks{These authors contributed equally to this work.}
\affiliation{Hefei National Laboratory for Physical Sciences at the Microscale and Department of Modern Physics, University of Science and Technology of China, Hefei, 230026, China.}
\affiliation{Synergetic Innovation Center of Quantum Information and Quantum Physics, University of Science and Technology of China, Hefei, 230026, China.}

\author{Jianpei Geng}
\thanks{These authors contributed equally to this work.}
\affiliation{Hefei National Laboratory for Physical Sciences at the Microscale and Department of Modern Physics, University of Science and Technology of China, Hefei, 230026, China.}

\author{Fazhan Shi}
\affiliation{Hefei National Laboratory for Physical Sciences at the Microscale and Department of Modern Physics, University of Science and Technology of China, Hefei, 230026, China.}
\affiliation{Synergetic Innovation Center of Quantum Information and Quantum Physics, University of Science and Technology of China, Hefei, 230026, China.}

\author{Ying Liu}
\author{Kebiao Xu}
\author{Wenchao Ma}
\author{Fei Kong}
\author{Zhen Jiang}
\author{Yang Wu}
\affiliation{Hefei National Laboratory for Physical Sciences at the Microscale and Department of Modern Physics, University of Science and Technology of China, Hefei, 230026, China.}

\author{Jiangfeng Du}
\email{djf@ustc.edu.cn}
\affiliation{Hefei National Laboratory for Physical Sciences at the Microscale and Department of Modern Physics, University of Science and Technology of China, Hefei, 230026, China.}
\affiliation{Synergetic Innovation Center of Quantum Information and Quantum Physics, University of Science and Technology of China, Hefei, 230026, China.}

\date{\today}

\begin{abstract}
Quantum computation provides great speedup over its classical counterpart for certain problems.
One of the key challenges for quantum computation is to realize precise control of the quantum system in the presence of noise.
Control of the spin-qubits in solids with the accuracy required by fault tolerant quantum computation under ambient conditions remains elusive.
 Here, we quantitatively characterize the source of noise during quantum gate operation and demonstrate strategies to suppress the effect of these. A universal set of logic gates in a nitrogen-vacancy center in diamond are reported with an average single-qubit gate fidelity of $0.999952$ and two-qubit gate fidelity of  $0.992$. These high control fidelities have been achieved at room temperature in naturally abundant $^{13}$C diamond via composite pulses and an optimised control method.
\end{abstract}

\maketitle

Quantum computations promise solutions of certain intractable problems in classical computations, such as quantum simulations \cite{QS,QS1}, prime factoring \cite{QFA1, QFA2}, and machine learning \cite{QAI1,QAI2}.
Recently, exciting progress towards spin-based quantum computation has been made with nitrogen-vacancy (NV) centers in diamond, which is a promising candidate for quantum computation under ambient conditions \cite{PNAS_DCRTQP}.
Long coherence time of NV centers has been achieved with dynamical decoupling technique \cite{Sci_UDDSSSSSB}.
Robust single-qubit and two-qubit gates have been accomplished \cite{ PRL_PQCIGODDS, PRL_IDCGSESD, Nat_DPQGHSSSR, Duan_2014}.
NV-NV entanglement has been realized showing the scalability of this quantum system \cite{NatCom_HFSEUOC, Nat_HESSQSTM}.
Quantum algorithms \cite{PRL_RTIDJASESD,Nat_DPQGHSSSR} and quantum error correction \cite{Nat_QECSSHSR, NatNano_UCECMQSRD} have been recently reported in NV centers.
Further improvement of NV center towards realistic quantum computation would require high fidelity quantum gates with errors below fault-tolerant threshold, which is proposed to be between $10^{-4}$ and $10^{-2}$ depending on the noise model and the computational overhead for realizing quantum gates \cite{ Nat_QCRND, PRA_2011, Science345_302}.
While fault-tolerant control fidelity has been reported very recently in superconducting qubits \cite{Nature_2014_Barends}, trapped ions \cite{PRL113_220501_Harty}, and  phosphorus doped in silicons \cite{NatNano9_981}, it is still of great challenge to achieve fault-tolerant fidelity under ambient conditions, which is the case for NV center in $^{13}$C naturally abundant diamond at room temperature.
It is because that the noise, not only stemming from interactions between the quantum system and the environment but also induced by imperfect manipulations, limits the fidelities of quantum gates.

In the following, a universal set of high-fidelity quantum gates at the threshold for fault-tolerant quantum computations in the NV center system are realized.
A novel composite pulse technique has been developed to suppress the noises during the single-qubit gates. We adopt the quantum optimal control method for designing the two-qubit gate to suppress the effects induced by both  the spin bath and the imperfect control field.
A novelly designed coplanar waveguide (CPW) with $15~$GHz bandwidth has been fabricated to minimize the effect due to the finite-bandwidth for the microwave pulses. Pulse fixing technique is utilized to correct the effect of the imperfect generation of the microwave pulses.
The average gate fidelity of single qubit is measured to be $0.999952(6)$ and the fidelity of the two-qubit controlled-NOT (CNOT) gate reaches $0.9920(1)$.  Thus we have successfully demonstrated a universal set of quantum gates with the fault-tolerant control fidelity in diamonds.

\begin{figure*}
\centering
\includegraphics[width=0.7\textwidth]{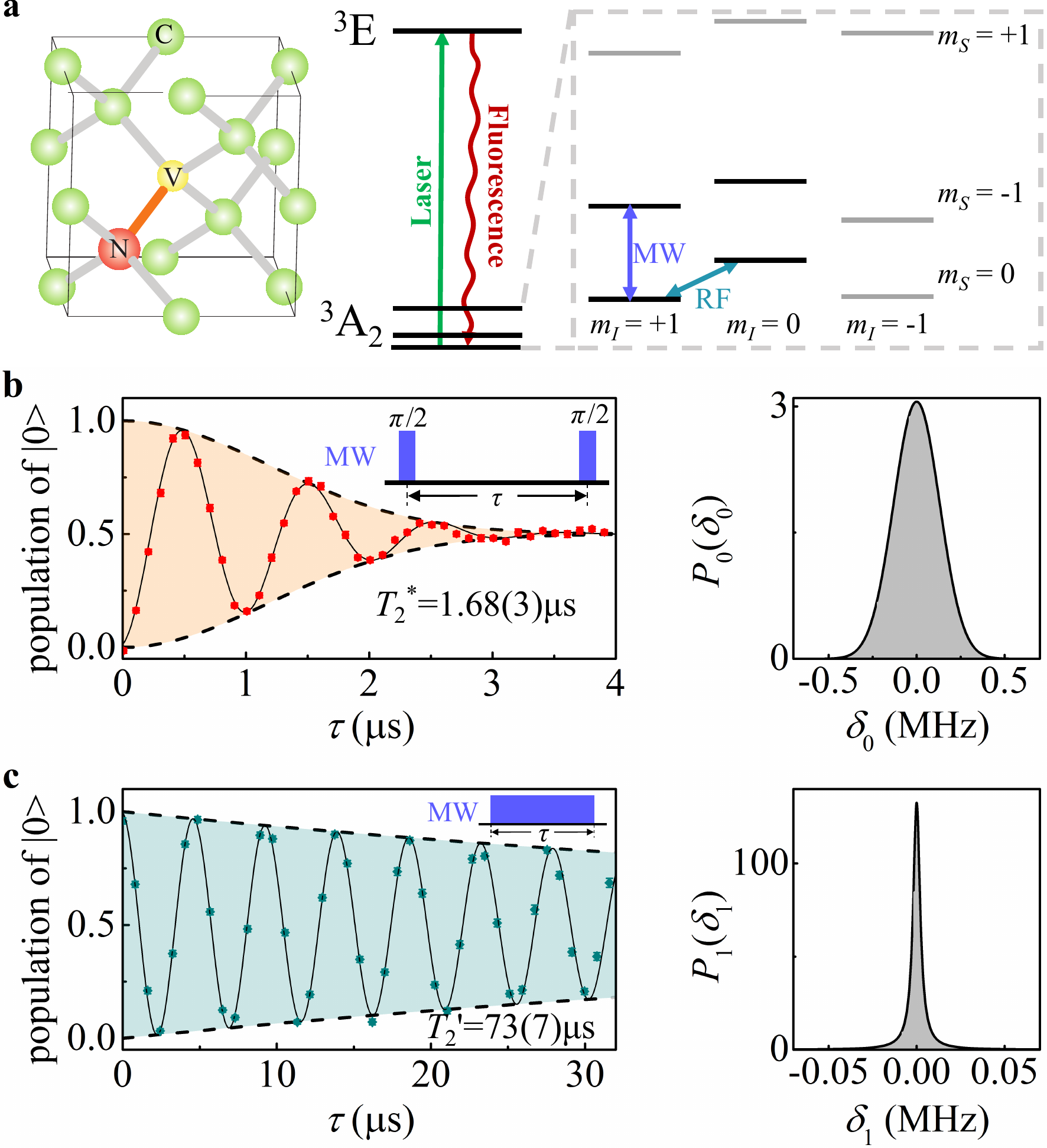}
\renewcommand{\figurename}{Figure}
    \caption{\textbf{Characterization of the noises in experiment.} $\textbf{a}$, Schematic atomic and energy levels of the NV center. Laser pulses are used for state initialization and readout. Microwave pulses (MW) are used for coherent control of the electron spin. Radiofrequency pulses (RF) are used for manipulating the nuclear spin. $\textbf{b}$, Result of the Ramsey experiment (inset, pulse sequence) for the electron spin. The solid black line in the left panel is fit to the experimental data (red circles) with $P_0(\delta_0)$ , which is the probability density distribution of $\delta_0$ depicted in the right panel. The decay time of FID is measured to be $T_2^* = 1.68 (3)~\mu$s.
    $\textbf{c}$,  Results of the nutation experiment (inset, pulse sequence) for the electron spin. The stepped MW pulse length is set to be $810~$ns. The decay time of the nutation is $T_2'= 73(7)~\mu$s. The solid black line in the left panel is fit to the experimental results (olive diamonds) with $P_1(\delta_1)$ , the probability density distribution of $\delta_1$ depicted in the right panel. The error bars on the data points are the standard deviations from the mean.
 }
    \label{Fig1}
\end{figure*}

\textbf{Results}

\textbf{NV center in diamond.} Fig. 1a depicts the NV center in diamond and the energy level structure. The NV center consists of a substitutional nitrogen atom with an adjacent vacancy cite (V) in the diamond crystal lattice. The ground state of NV center is an electron spin triplet state $^3\textrm{A}_2$, with three sublevels $|m_S=0\rangle$ and $|m_S=\pm1\rangle$. The intrinsic nitrogen nuclear spin results further splitting of the energy levels due to the hyperfine coupling.
In the experiment presented here, the two-qubit system comprises the electron spin qubit and $^{14}$N nuclear spin qubit. Electron (nuclear) spin states $|m_{S} = 0\rangle$ and $|m_{S} = -1\rangle$ ($|m_{I}\rangle= 0\rangle$ and $|m_{I} = +1\rangle$ ) are encoded as the electron (nuclear) spin qubit ( Fig. 1a).
When a $532~$nm green laser pulse is applied to NV center, the electron spin will be excited to the $^3\textrm{E}$ state, and then fluorescence emission can be measured.
The optical transitions are used to initialize and read out the state of the electron spin. The polarization of the electron spin is measured to be $0.83(2)$  (see Supplementary Note 8 and Supplementary Fig. 9). A magnetic field of $513~$Gauss is applied along the NV symmetry axis ([1 1 1] crystal axis) to split the $|m_S=\pm1\rangle$ energy levels, and to polarize the nitrogen nuclear spin (Supplementary Note 5).

\textbf{Realization of high fidelity single-qubit gates.} The high-fidelity single-qubit gate is demonstrated on the electron spin qubit. The state evolution of the electron spin qubit is governed  by the Hamiltonian $H_{\textrm{ideal}}=2\pi\omega_1(\cos\phi S_x+\sin\phi S_y)$ in the rotating frame, where $\phi$ is the microwave phase and $\omega_1$ is the Rabi frequency.
Unfortunately, this is only the ideal case.
In practice, the Hamiltonian is $H_{\textrm{prac}}=2\pi\delta_0S_z+2\pi(\omega_1+\delta_1)[\cos (\phi + \delta\phi)S_x+\sin (\phi + \delta\phi) S_y]$, where $\delta_0$, $\delta_1$, and $\delta\phi$ lead to gate error.
The $\delta_0$ mainly comes from the Overhauser field (due to the interaction with the nuclear spin bath), the magnetic field fluctuation, and the instability of the microwave frequency.
The $\delta_1$ partly comes from the control field error due to the static fluctuation of the microwave power. The distortion of the pulse due to the finite bandwidth also contributes to the $\delta_1$ .
The $\delta\phi$ describes the distortion of the microwave phase due to the imperfect microwave generation.
We take the distortion of the amplitude and the phase of the microwave pulse as the systematic errors, which have unchanged deviations from the ideal cases during each experimental scan. The other errors can be treated as the fluctuation of the experimental parameters. All of these errors limit the fidelity of quantum gates.

We first quantitatively characterize the distributions of errors due to the fluctuation of the experimental parameters. Since the timescale of variation of $\delta_0$ and $\delta_1$ is much longer than that of the quantum gates, $\delta_0$ and $\delta_1$ can be taken as quasi-static random constants, and the practical Hamiltonian is rewritten as $H_{\textrm{prac}}=2\pi\delta_0S_z+2\pi(\omega_1+\delta_1)S_x$.
The fluctuations of $\delta_0$ and $\delta_1$ lead to the decoherence of the qubit during quantum gates when averaged over repeated experiments. Fig. 1b and 1c show the experimental distributions of $\delta_0$ and $\delta_1$ derived by Free Induction Decay (FID) and nutation experiments, respectively.
The distribution of $\delta_0$, i.e. $P_0 (\delta_0)$, is obtained from the FID signal via the Ramsey experiment, which is shown in the left panel of Fig. 1b.
The oscillation of the FID signal  is due to the detuning of the microwave frequency.
We assume that $\delta_0$ satisfies a Gaussian distribution $P_0(\delta_0)=\exp (-\delta_0^2 /2\sigma^2)/(\sigma\sqrt{2\pi})$, where $\sigma$ stands for the  standard deviation of the distribution.
The fitting of the FID data is based on distribution $P_0(\delta_0)$, with parameter $\sigma$  optimized to achieve best agreement between the fitted and experimental data.
The fitted data with best agreement is shown as the solid curve in Fig. 1b, which gives $\sigma=0.131(5)~$MHz, with the probability density distribution $P_0(\delta_0)$ depicted in the right panel of Fig. 1b.
The distribution of $\delta_1$, i.e. $P_1 (\delta_1)$, obtained via nutation experiments, is shown in Fig. 1c.  The Rabi frequency is set to be $\omega_1=10~$MHz.
The $\delta_1$ satisfies a Lorentzian distribution of $P_1(\delta_1)=\gamma/ \pi(\delta_1^2+\gamma^2)$, where $\gamma$ is the half-width at half-maximum of the distribution.
The distribution $P_1(\delta_1)$ together with $P_0(\delta_0)$ obtained from the FID, is used to fit the nutation experiment. Best agreement between the fitted and experimental data is achieved with $\gamma=0.0024(4)~$MHz. The fitting result is the solid curve in the left panel of Fig. 1c. The probability density distribution $P_1(\delta_1)$ is depicted in the right panel of Fig. 1c.

\begin{figure*}
\centering
\includegraphics[width=0.7\textwidth]{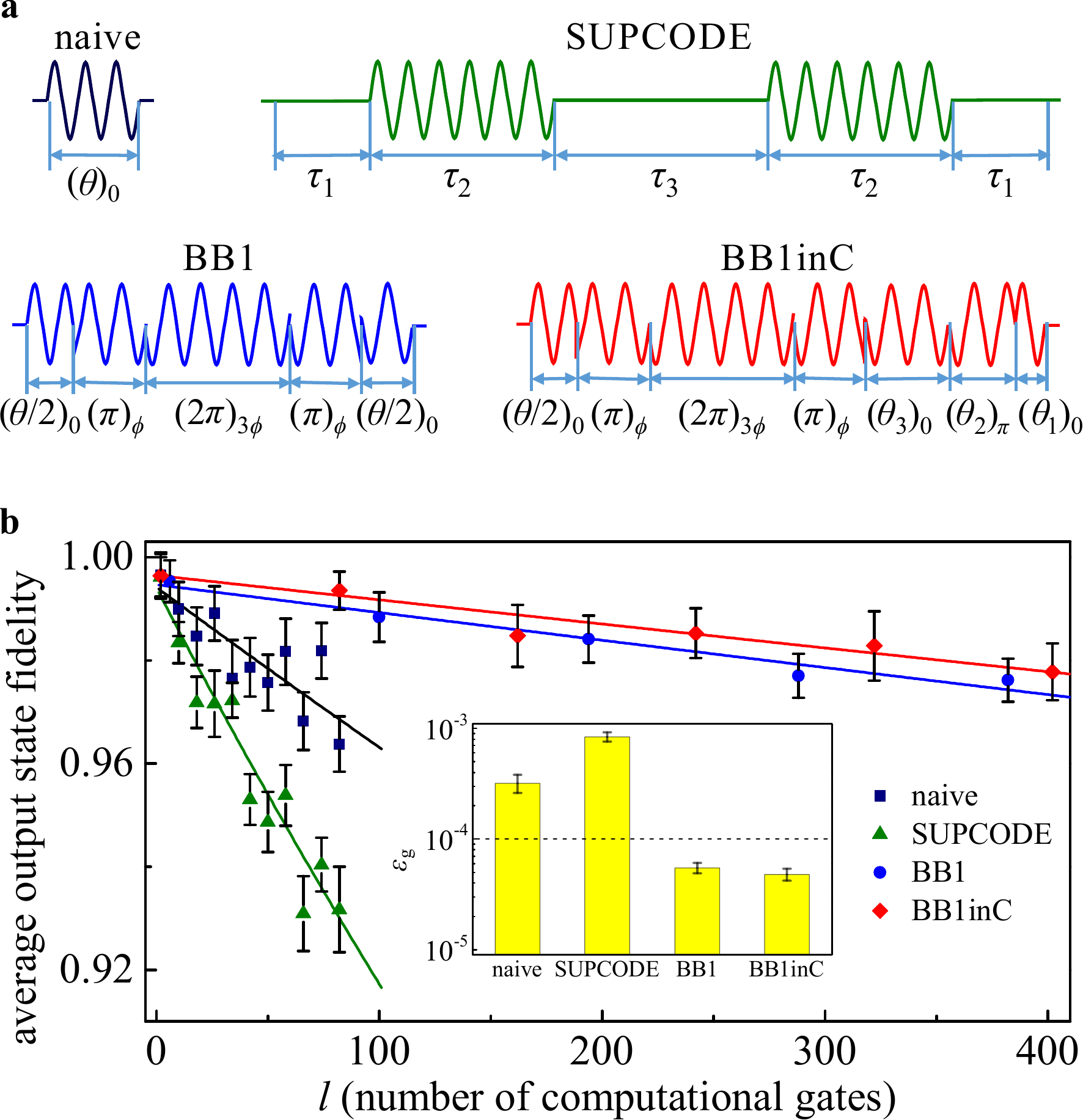}
\renewcommand{\figurename}{Figure}
    \caption{\textbf{Characterization of the performance of single-qubit gates.} $\textbf{a}$,  Pulse sequences corresponding to naive pulse, SUPCODE pulse, BB1 pulse and BB1inC pulse. The time durations $\tau_i$ ($i$ = 1, 2 and 3) of SUPCODE is included in Supplementary Note 3. $\textbf{b}$,  Results of randomized benchmarking.
    Red diamonds, blue circles, navy blue squares, and green triangles represent measured single-qubit fidelity of the output state from each of the individual sequences of gates.
   Solid lines are fits of equation (\ref{eq1}) with averaging each over all the randomization pulse sequences. The average gate fidelity for the naive, five-piece SUPCODE,  BB1 and BB1inC pulses are 0.99968(6), 0.99916(8),  0.999945(6), and 0.999952(6) respectively. The inset shows the average error per gates ($\varepsilon_\textrm{g}$) of the pulses. The error bars on the data points are the standard deviations from the mean, and those of $\varepsilon_\textrm{g}$ in inset are given by errors of the fit of the randomized benchmarking data.
 }
    \label{Fig2}
\end{figure*}

Since distributions of $\delta_0$ and $\delta_1$ have been quantitatively characterized, we now demonstrate the realization of high-fidelity gates with a novel composite pulse, which is designed to  suppress these noises simultaneously.
To form the new composite pulse, we replace one of the constituent elementary pulses of CORPSE (compensation for off-resonance with a pulse sequence) by BB1 (broadband number 1) sequence. This new composite pulse is named as BB1inC for short. Theoretical calculation shows that BB1inC is robust against both errors of $\delta_0$ and
$\delta_1$ (Supplementary Note 3 and Supplementary Fig. 1). We experimentally compare the performance of this new pulse sequence with three other types of pulse sequences, which are naive (rectangular) pulse, SUPCODE (soft uniaxial positive control for orthogonal drift error) \cite{NatCommun_CPRUCSTQ} pulse, and BB1 pulse \cite{JMR_BNPCPUANMRE}.
The rotation of an angle $\theta$ around the axis in the equatorial plane with azimuth $\phi$ is denoted by $(\theta)_{\phi}$.
The naive pulse is very sensitive to the errors $\delta_0$ and $\delta_1$, with leading orders of both errors preserved in the evolution operator (corresponding to second orders in gate fidelity).
The SUPDODE pulse, a type of dynamically corrected gate, has been proposed to suppress the dephasing noise during quantum gates.
The five-piece SUPCODE pulse has been used here, where the waiting time $\tau_{1(3)}$ and the pulse duration $\tau_2$ satisfy the specific requirement (see Supplementary Note 3). Under the five-piece SUPCODE pulse, up to second order of $\delta_0$ can be canceled (corresponding to sixth order preserved in gate fidelity).
This has recently been demonstrated in NV centers \cite{PRL_IDCGSESD} with control field $\omega_1$ of $1~$MHz.
However, if we increase the control field to shorten the gate time, the fluctuation of the control field will dominate the decrease of the gate fidelity.
A composite pulse, named BB1, which is shown to be robust against the $\delta_1$, can be applied to suppress the fluctuation of the control field. This pulse sequence is
$(\theta / 2)_0-(\pi)_{\phi}-(2\pi)_{3\phi}-(\pi)_\phi-(\theta / 2)_0$, with $\phi=\arccos(-\theta / 4\pi)$. The error induced by $\delta_1$ is inhibited up to second order in the evolution operator.
The sequence of BB1inC, which consists of seven pulses, is shown in Fig. 2a.
The BB1inC sequence is depicted as $(\theta/2)_0-(\pi)_{\phi}-(2\pi)_{3\phi}-(\pi)_{\phi}-(\theta_3)_0-(\theta_2)_{\pi}-(\theta_1)_0$, where $\phi=\arccos(-\theta/4\pi)$, $\theta_1=\theta/2-\arcsin[\sin(\theta/2)/2]$, $\theta_2=2\pi-2\arcsin[\sin(\theta/2)/2]$, and $\theta_3=2\pi-\arcsin[\sin(\theta/2)/2]$. Leading orders of both the $\delta_0$ and $\delta_1$ errors are canceled in the evolution operator, so BB1inC can suppress both errors simultaneously.

 The average gate fidelity \cite{PLA_SFAGFQDO} between a quantum operation $\xi$ and a target unitary quantum gate $U$ is defined as
\begin{equation}
\label{Fave}
F_{\textrm{a}}(\xi, U) =  \int d\psi \langle\psi|U^\dag \xi(|\psi\rangle\langle\psi|)U|\psi\rangle,
\end{equation}
where the integral is over the uniform measure $d\psi$ on the Hilbert space of the system. To estimate the fidelity by equation (\ref{Fave}), it is necessary to  characterize the quantum operation $\xi$ with comprehensive knowledge about the Hamiltonian of the quantum system,  and the Hamiltonian of the  control together with the errors during the gate operation.
We numerically calculate the gate fidelities (Supplementary Note 2) for these pulse sequences with both $\delta_0$ and $\delta_1$ . The BB1inC pulse presents a desirable region with fidelity higher than $0.9999$ (Supplementary Note 3 and Supplementary Fig. 1). The control field strength is set to $\omega_1 = 10~$MHz for experimental realization of high-fidelity single-qubit gates, so that the error due to $\delta_0$ and $\delta_1$ can be greatly reduced.

Though composite pulses provide possibility to realize high fidelity gates in solid-state system, the performance of the composite pulses is limited by the systematic errors, which are usually not taken into account in the composite pulse designing.
Herein, we presents methods to correct the systematic errors. The distortion of the microwave phase is corrected by pulse fixing technique, which is previously developed in liquid NMR (nuclear magnetic resonance) \cite{PRL107_170503}.
The phase of the microwave pulses was recorded by an oscilloscope, and then the deviation of the phase from the ideal case was fed back to the arbitrary waveform generator so that this distortion can be minimized (see Supplementary Note 6 and Supplementary Fig. 5).
The distortion of the amplitude is primarily due to the limited bandwidth of the microwave fed structure.
We designed and fabricated an ultra-broadband CPW with a bandwidth up to $15~$GHz (see Supplementary Note 5 and Supplementary Fig. 4).
Then the effect of the finite bandwidth is found to be negligible.
The reflection between the microwave components also contributes to the imperfection of the microwave pulses.
This effect was diminished by inserting  proper attenuators between the microwave components (Supplementary Note 6 and Supplementary Fig. 6). The amplitude distortion is further corrected with pulse fixing (Supplementary Note 6 and Supplementary Fig. 7). After all these procedures, the microwave pulses can be fed to the electron spin with almost perfect pulse shapes . The detailed procedures for correcting the systematic errors are included in the Supplementary Note 6.

\begin{figure*}[htb]
\centering
\includegraphics[width=0.7\textwidth]{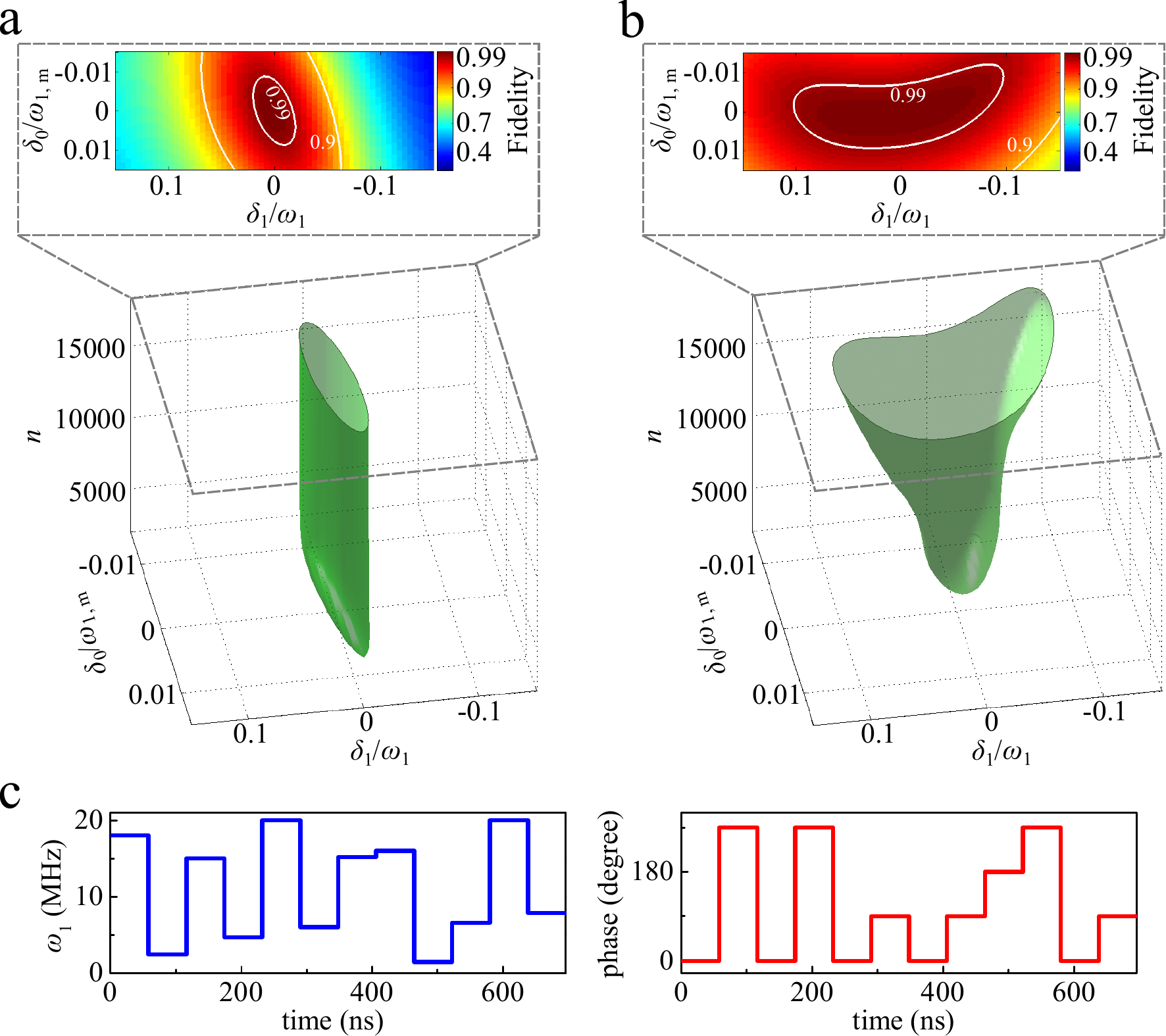}
\renewcommand{\figurename}{Figure}
    \caption{\textbf{Optimization of the control of the two-qubit system to achieve high fidelity control in the presence of noises.} $\textbf{a}$ and $\textbf{b}$, Comparison of the original GRAPE (a) with the modified one (b) in the presence of the noises $\delta_0$ and $\delta_1$. Here $\omega_{1\textrm{,m}}$ stands for the maximal strength of the control field in the pulse sequence, and $n$ stands for the number of iterations in the algorithms.
     The upper figures of (a) and (b) are the fidelity contour maps of the two GRAPE methods. The area with a control fidelity higher than 0.99, is larger for the modified method than the original one. $\textbf{c}$, The schematic diagram of the optimal control pulses used in the experiment, with the left and right being the amplitude and phase sequences, respectively.
 }\label{Fig3}
\end{figure*}

To experimentally quantify the performance of the single-qubit gates, we utilize the randomized bechmarking (RB) method \cite{PRA_RBQG}. In the RB experiment, quantum gates are evaluated by measuring the performance when random sequences of the gates are applied. The error due to the imperfect measurement and state preparation can be separated, and the gate fidelity can be determined precisely.
When the number of randomized gates implemented in a sequence is increased, the accumulated gate error reduces the measured single-qubit fidelity $F$ of the output state, which is defined as the overlap of the ideal and the measured states.
The average output state fidelity $\overline{F}$ \cite{PRA_RBQG} can be written as
\begin{equation}
\overline{F}= 1/2  + 1/2 (1-d_{\textrm{if}})(1-2\varepsilon_\textrm{g})^{l},
\label{eq1}
\end{equation}
where $d_{\textrm{if}}$ is due to the imperfection of the state initialization and readout, $\varepsilon_\textrm{g}$ is the average error per gate, and $l$ is the number of randomized quantum gates.
The average gate fidelity ($F_{\textrm{avg}} = 1 -\varepsilon_\textrm{g} $) derived from this method is resilient to the state preparation and measurement errors.
The detailed procedure of RB is included in the Supplementary Note 9.
Fig. 2b shows the result of the randomized bechmarking, where the control field $\omega_1$ is set to be $10~$MHz.
The average gate fidelity of naive pulses is found to be 0.99968(6).
The five-piece SUPCODE pulses provide an average gate fidelity of 0.99916(8) which is lower than that of the naive one.
This is because the fluctuation of control field dominates the decay of the fidelity when $\omega_1 = 10~$MHz.
The average gate fidelity is greatly improved when BB1 pulses are applied. The extracted average gate fidelity for BB1 pulses is $0.999945(6)$. The highest fidelity is achieved when BB1inC pulses are used. The average gate fidelity for BB1inC is $0.999952(6)$, providing an error per gate below $10^{-4}$ which is about one order of magnitude lower than that of naive pulses (Supplementary Table 1).

\textbf{Realization of high fidelity CNOT gate.} Two-qubit CNOT gate together with the single-qubit gates provide a universal set of quantum gates. Two-qubit gates have been demonstrated in NV centers \cite{Nat_DPQGHSSSR,Duan_2014}.
However, realization of the two-qubit gates with high fidelity to meet the requirement of the fault-tolerant quantum computation is still of great challenge.
The control qubit is the nuclear spin qubit and the target qubit is the electron spin qubit. The detailed Hamiltonian is included in Supplementary Note 1. The CNOT gate is designed to flip the electron spin qubit iff the nuclear spin qubit is in the state $|m_{I} = 1\rangle$.
Microwave (MW) and radio-frequency (RF) pulses are used to manipulate this two-qubit system.

To achieve high fidelity, we have improved a type of quantum optimal control method, named GRAPE (gradient ascent pulse engineering) \cite{GRAPE}, to design the CNOT gate which is robust against both $\delta_0$ and $\delta_1$.
GRAPE has recently been used to realize the quantum error correction \cite{Nat_QECSSHSR} and a high-fidelity entanglement \cite{NatCom_HFSEUOC} in diamond.
Because the nuclear spin is insensitive to the noise (such as fluctuations of the external magnetic field and the control RF field) and the CNOT gate designed by the quantum optimal control method consists of microwave pulses only, we can take into account the noise felt by the electron spin in the optimization procedures.
We modified the GRAPE algorithm to design target gates which are robust against both the errors $\delta_0$ and $\delta_1$ (see Supplementary Note 4 and Supplementary Fig. 2).
Fig. 3 shows the optimization of microwave pulse parameters to achieve the high control fidelity.
Fig. 3a (b) shows the optimizing procedure of original (modified) GRAPE algorithm.
The final achieved high fidelity regions with the two GRAPE algorithms are compared in the upper parts of Fig. 3a  and 3b.
It is clear that the modified method presents much more robustness against both $\delta_1$ and $\delta_0$ than the original one.
The left and right panels of Fig. 3c show the amplitude and phase sequences after the improved optimization, respectively.
The total length of the sequence for one CNOT gate is $696~$ns, which consists of twelve pulses of equal length with different amplitudes and phases. The theoretical fidelity of the CNOT gate via this pulse sequence is estimated to be $0.9995$ in the absence of $\delta_0$ and $\delta_1$. If the noises $\delta_0$ and $\delta_1$  are taken into account, the average gate fidelity of the designed CNOT gate is estimated to be $0.9927$. The detailed pulse optimization and fidelity estimation are included in the  Supplementary Note 4.

Fig. 4 presents the experimental realization of the CNOT gate. The two-qubit system is prepared to the state $|01\rangle $ and  $|00\rangle$ in Fig. 4a and 4c, respectively (The state $|m_{S}=0, m_{I}=1\rangle$ is denoted as $|01\rangle $ and the state $|m_{S}=0,m_{I} = 0\rangle $ as $|00\rangle $ hereafter).
The black (red) symbols with error bars in  Fig. 4a and 4c are the experimental FID signal via the Ramsey experiment without (with) the CNOT gate.
Fig. 4b and 4d are the fast Fourier transformation of the FID signals, so that
the result of the CNOT gate can be observed in the frequency domain.
The two peaks correspond to the nuclear spin qubit states $|m_{I} = 0\rangle$ and $|m_{I} = 1\rangle$. The distance between  the two peaks is due to the hyperfine coupling.
It is clear that when the nuclear spin qubit is of state $|m_{I} = 1\rangle$, the electron spin qubit is rotated by $\pi$ ( Fig. 4a and 4b). If the nuclear spin qubit is of state $|m_{I} = 0\rangle$, the electron spin qubit remains unchanged (see Fig. 4c and 4d).

Two-qubit randomized benchmarking, which can be used to measure the fidelity of two-qubit gate, requires operations on both qubits \cite{Nature_2014_Barends}. In the hybrid system composed of electron and nuclear spins, single-qubit gates of the nuclear spin cost much longer time than that of the electron spin qubit.
The typical operation time on the nuclear spin qubit (about $50~\mu$s for a $\pi$ rotation) is much longer than the dephasing time $T_2^*$ of electron spin qubit.
The decoherence effect on the electron spin during the nuclear spin operation in the two-qubit RB experiments will  dominate the fidelity decay in randomized benchmarking, and the gate fidelity of CNOT
can not be precisely determined in this way. On the other hand, though process tomography provides a full characterization of CNOT gate, the measured gate fidelity with this method is sensitive to errors in state preparation and measurement. In this scenario, repeated application of CNOT gates on the system and recording the dynamics of the quantum state \cite{PRL109_060501_Chow} will be a good choice to estimate the gate fidelity.

Fig. 4e shows the population of $|01\rangle$, $P_{|01\rangle}$ ,  after repeated applications of CNOT gates on the input state, which is generated by a selective $\pi/2$ RF pulse. The inset of Fig. 4e shows the pulse sequence, where the number of repeated CNOT, $N$, is even. When $N$ is increased, the error of the CNOT gate will accumulate and $P_{|01\rangle}$ will decay.
A wealth of information can be obtained by studying the state dynamics under repeated applications of CNOT gates.
In ref. \onlinecite{PRL109_060501_Chow}, the dynamics of states obey a simple exponential decay and the gate fidelity can be extracted with twelve applications of CNOT gates. However, when more CNOT gates (up to $192$ in this work) are applied, we find that the decay is not exponential.
So the gate fidelity cannot be simply extracted from the decay as presented in the ref. \onlinecite{PRL109_060501_Chow}.
 In Fig. 4e, the dynamics of the state exhibits oscillatory attenuation as $N$ increases.
The oscillation results from the deviation of the realistic evolution from the ideal CNOT operation.
For example, the optimization procedure adopts the hyperfine coupling $A = -2.16~$MHz, which may differ slightly from the experimental value $A_{\textrm{exp}}$ by $\delta A = A_{\textrm{exp}}-A$.
The frequency of the microwave will not be equal to the resonant frequency exactly, which induces an off-resonance term ($\delta\Omega$) in the practical Hamiltonian.
The decay of the $P_{|01\rangle}$  is mainly due to the static fluctuation of $\delta_1$ and $\delta_0$, which can be quantitatively characterized by the FID and the nutation experiment, respectively.
The values of $\delta A$ and $\delta\Omega$ were derived to be $0.008~$MHz and $0.068~$MHz, by fitting the experimental data (Supplementary Note 9 and Supplementary Fig. 10). The fitting result shown as the solid blue line in Fig. 4e agrees with the experimental data.
Since the comprehensive information on the Hamiltonian of the quantum system and the control field together with the errors is ready, the gate fidelity of CNOT can be directly obtained according to equation (\ref{Fave}) with $0.9920(1)$ (see Supplementary Note 9).

\begin{figure*}
\centering
\includegraphics[width=0.7\textwidth]{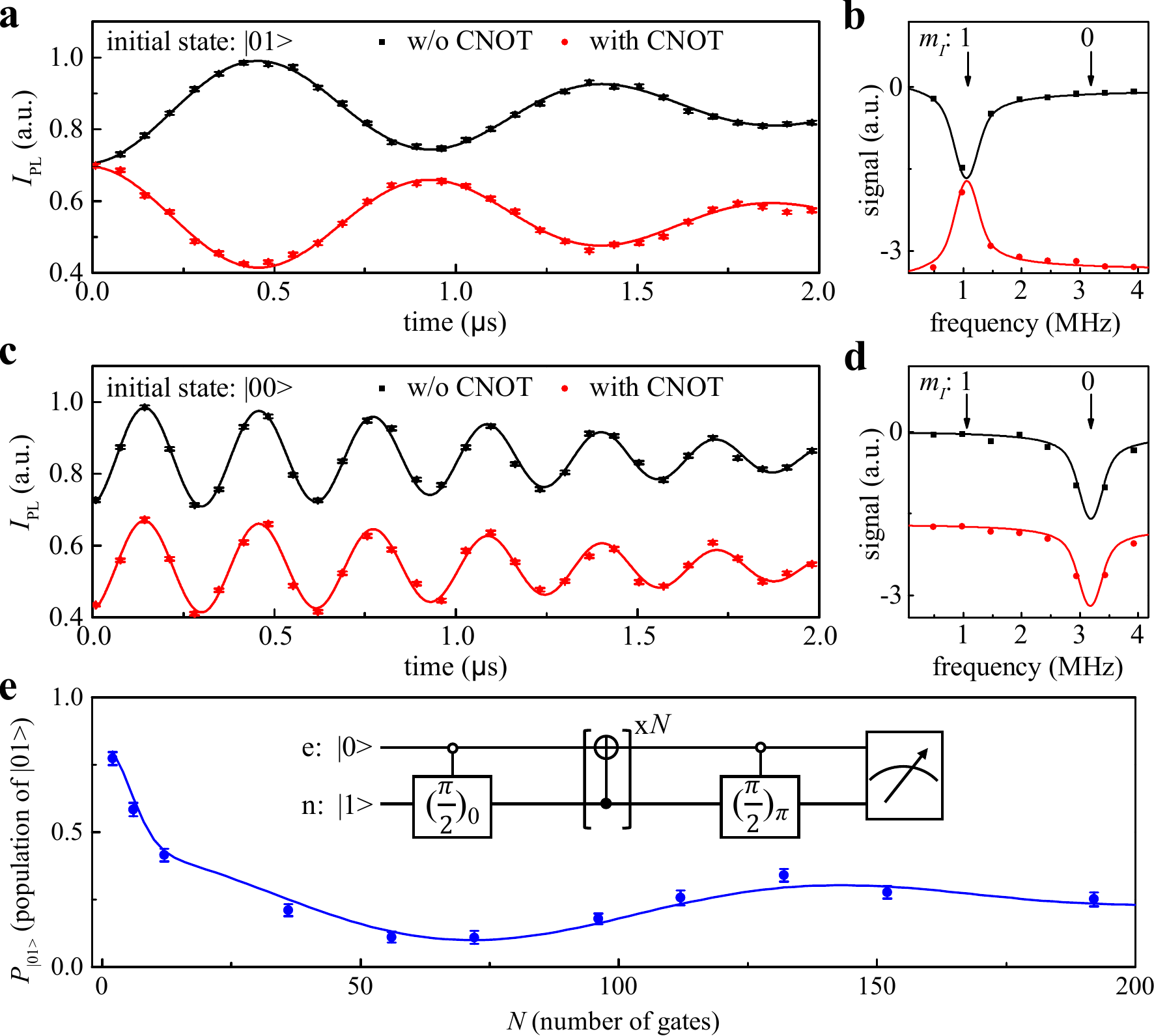}
\renewcommand{\figurename}{Figure}
    \caption{\textbf{Characterization of the performance of CNOT gate.} $\textbf{a}$  and $\textbf{c}$  are experimental data and the fittings of FID (black : without applying the CNOT gate, red: with applying the CNOT gate), when the initial states are $|01\rangle$ and  $|00\rangle$, respectively. Lines are fits to the experimental data (symbols with error bars). $\textbf{b}$ and $\textbf{d}$ are the fast Fourier transform results of FID signals. $\textbf{e}$,  The probability in the state $|01\rangle$ ($P_{|01\rangle}$) after applying CNOT gates repeatedly, with the sequence shown in the inset.
    The fitting (blue line) agrees with the experimental data (blue circles with error bars). The error bars on the data points are the standard deviations from the mean. The detailed procedure of normalization to obtain $P_{|01\rangle}$ is included in Supplementary Note 7 and Supplementary Fig. 8.
 }
    \label{Fig4}
\end{figure*}

\textbf{Discussion}

To sum up, we have achieved fault-tolerant control fidelity for a universal set of quantum gates in diamond with $^{13}$C of natural abundance under ambient conditions.
Several errors which limit the fidelity of quantum gates have been quantitatively characterized and effectively suppressed. With new composite pulses, we have realized single-qubit gate with fidelity up to $0.999952$.
A modified optimal control method has been developed to design the control pulse for the CNOT gate, which achieves a gate fidelity of $0.992$. To the best of our knowledge, our results stands for the state of art high-fidelity control of solid-state spins under ambient conditions.
Our method can not only be used to realize high fidelity CNOT gate in the system consisting of on the electronic spin and the host nitrogen nuclear spin in NV,  but also can be  applied  for the coupled NV-NV systems, which provide possibilities for future scalable architectures. The noises which limit the fidelity of the two-qubit gate in NV-NV system can also be greatly suppressed and high-fidelity gates are available by our method (Supplementary Note 10 and Supplementary Fig. 11).
The methods presented here to achieve high control fidelity can also be applied to other quantum systems, such as quantum dots, phosphorus doped in silicon, and trapped ions.

\textbf{Methods}

\textbf{Experiment setup.} The NV center in $[100]$ face bulk diamond was mounted on a typical optically detected magnetic resonance (ODMR) confocal setup which was synchronized with the microwave bridge by a multichannel pulse generator (Spincore, PBESR-PRO-500).  The nitrogen concentration was less than 5 ppb and the abundance of $^{13}$C was at the natural level of 1.1\%. The 532 nm green laser for pumping and phonon sideband fluorescence ( 650-800 nm) went through the same oil objective (Olympus, PLAPON 60XO, NA 1.42). In order to preserve the NV center's longitudinal relaxation time $T_1$  from laser leakage effects, the pump beam was passed twice through an acousto-optic modulator (ISOMET, power leakage ratio $\sim$1/1000) before it went into the objective.
We created a solid immersion lens (SIL) in the diamond around an NV center (see Supplementary Note 5 and Supplementary Fig. 3). The SIL increases the PL rate to about $400~$kcounts s$^{-1}$.
The fluorescence intensity was collected by an avalanche photodiodes (Perkin Elmer, SPCM-AQRH-14) with a counter card. The adjustable external magnetic field, created by the permanent magnets, was aligned by monitoring the variation of fluorescence counts. Microwave and radio frequency pulses were carried by ultra-broadband coplanar waveguide with $15~$GHz bandwidth.

{\bf Acknowledgements}\\
We are grateful to C. K. Duan, X. F. Zhai, and Z. K. Li for valuable discussion.
This work was supported by the National Key Basic Research Program of China (Grant No.\ 2013CB921800),
the National Natural Science Foundation of China (Grant No.\ 11227901, No.\ 31470835, and No.\ 11275183), and the Strategic Priority Research Program (B) of the CAS (Grant No. XDB01030400).  F. S. and X. R. thank support of the Youth Innovation Promotion Association of Chinese Academy of Sciences.

{\bf Author Contributions}\\
J. D. proposed the idea. J.D. and X. R. designed the experiments. J. G., Y. L., W. M., K. X., and F. S. performed the experiments. J. G., Y. W., and X. R. carried out the calculations.  Z. J. and X. R. designed the CPW.  X. R., J. G., and J. D. wrote the paper. All authors analysed the data, discussed the results, and commented on the manuscript.

{\bf Competing financial interests}\\
The authors declare no competing financial interests.

\section{Supplementary Figures}

\begin{figure*}[h]
\centering
\includegraphics[width=0.8\textwidth]{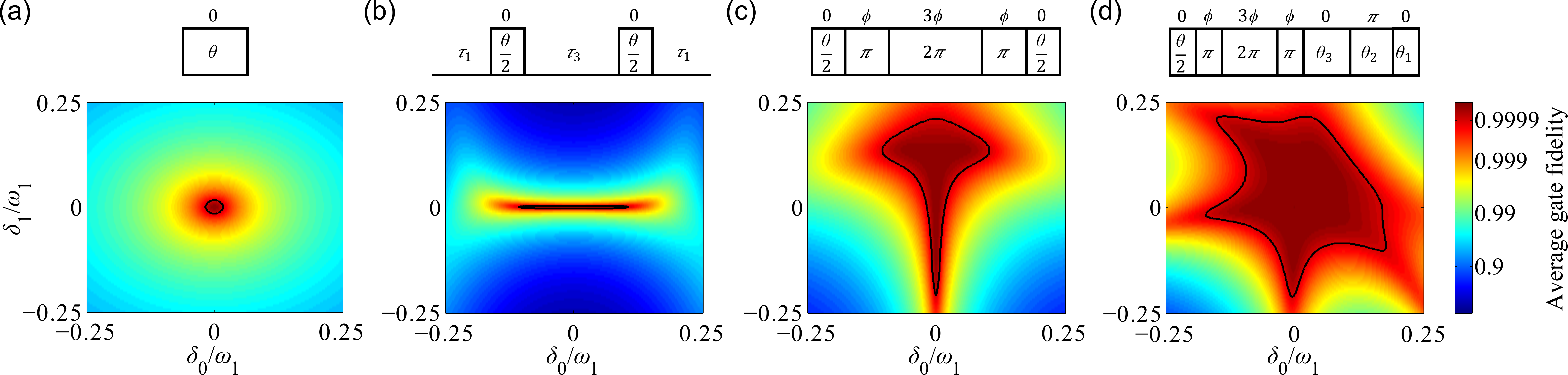}
\renewcommand{\figurename}{Supplementary Figure}
\caption{\label{Pulsedata} \textbf{Pulse sequences and contour plot of average gate fidelity of single-qubit gate.} The upper panels show the pulse sequences of $(\theta)_0$ with (a) rectangular naive, (b) five-piece SUPCODE, (c) BB1, and (d) BB1inC pulses, where the phase and duration of each piece of pulse is depicted. The lower panels show the average gate fidelity of $(\pi/2)_0$ with the errors $\delta_0$ and $\delta_1$ when (a) rectangular naive, (b) five-piece SUPCODE, (c) BB1, and (d) BB1inC pulses are applied. The regions of fidelity larger than 0.9999 are surrounded with black curves for clarity.}
\end{figure*}


\begin{figure*}[h]
\centering
\includegraphics[width=0.8\textwidth]{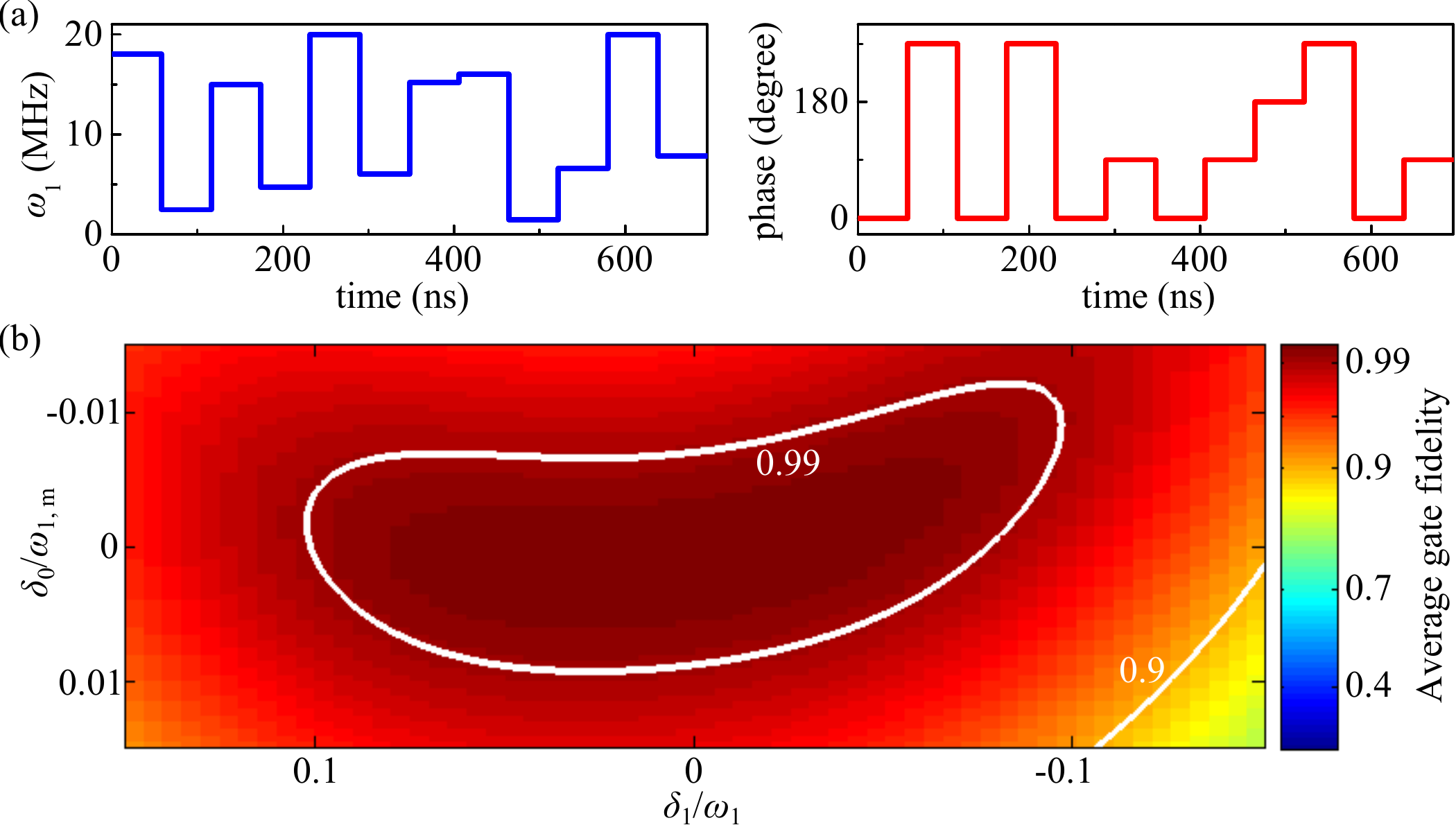}
\renewcommand{\figurename}{Supplementary Figure}
    \caption{\textbf{Designed pulse sequence for CNOT and average gate fidelity of the sequence.} (a) Amplitude (left panel) and phase (right panel) of the designed pulse sequence. (b) Calculated average gate fidelity of the sequence under the errors $\delta_0$ and $\delta_1$.
 }
    \label{GRAPEseqFave}
\end{figure*}

\begin{figure*}[h]
\centering
\includegraphics[width=0.8\textwidth]{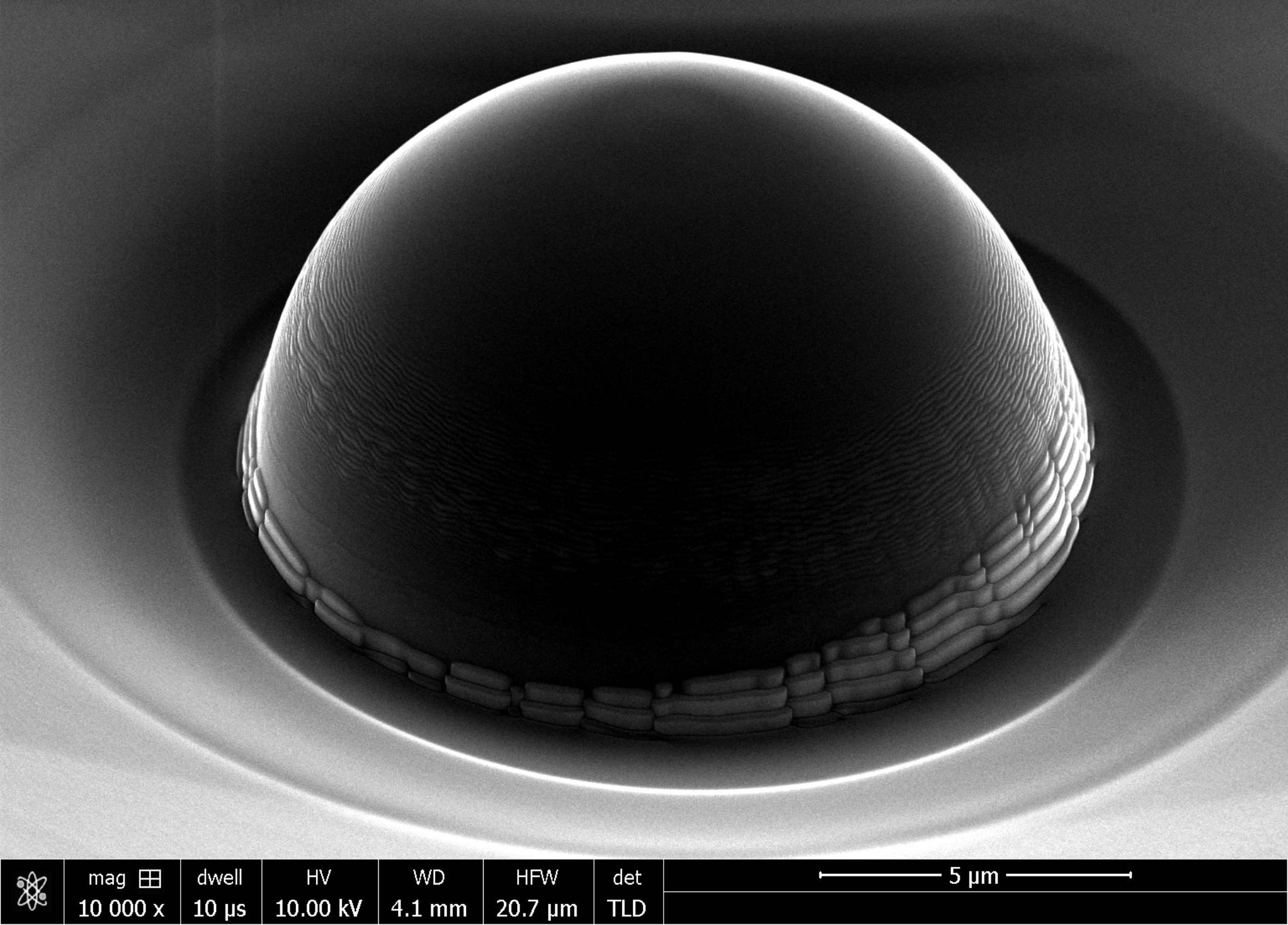}
\renewcommand{\figurename}{Supplementary Figure}
    \caption{\label{SIL} \textbf{Image of the SIL in diamond.}
 }
\end{figure*}

\begin{figure*}[h]
\centering
\includegraphics[width=0.8\textwidth]{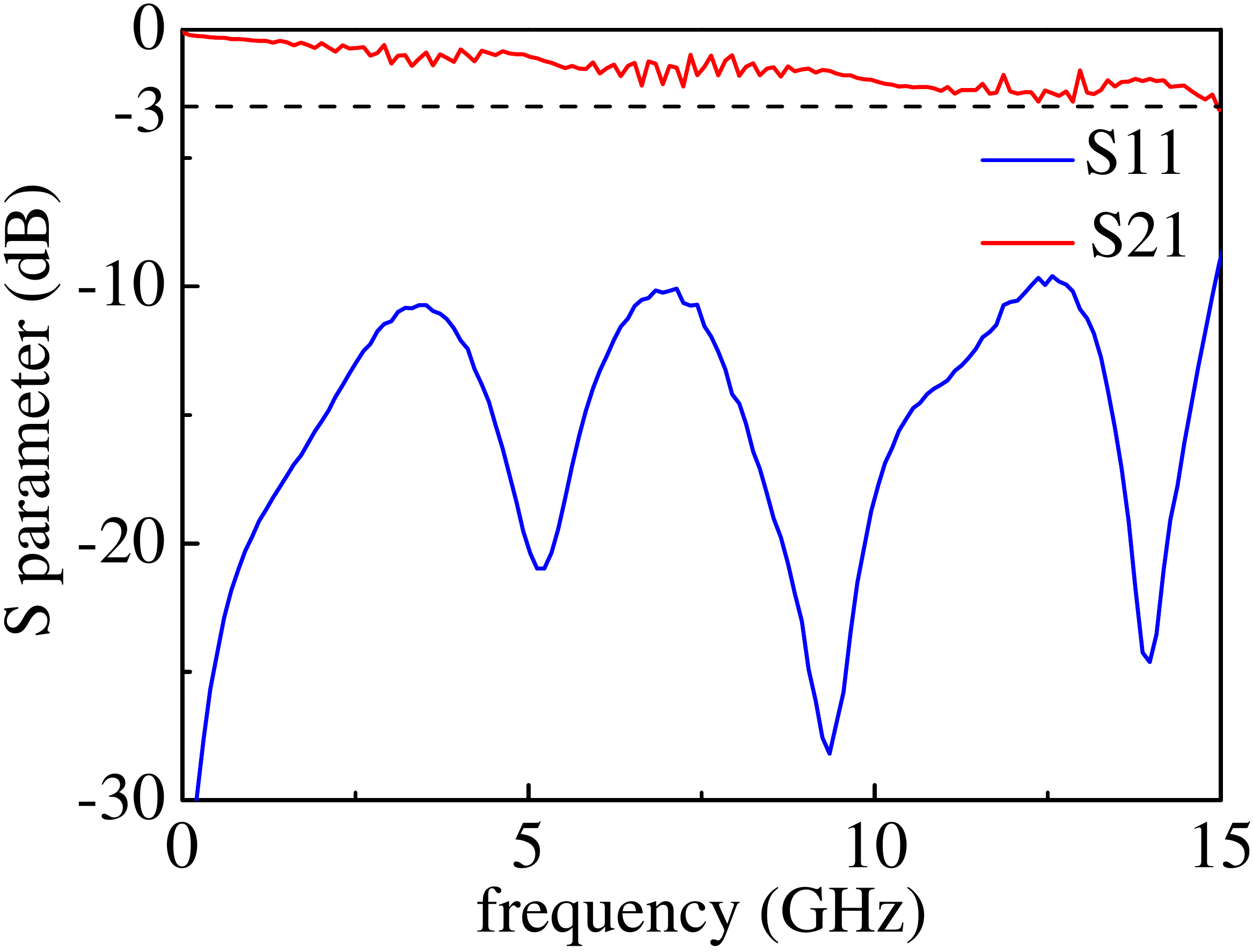}
\renewcommand{\figurename}{Supplementary Figure}
\caption{\label{CPWpara} \textbf{Scattering parameters of the CPW.}}
\end{figure*}


\begin{figure*}[h]
\centering
\includegraphics[width=0.8\textwidth]{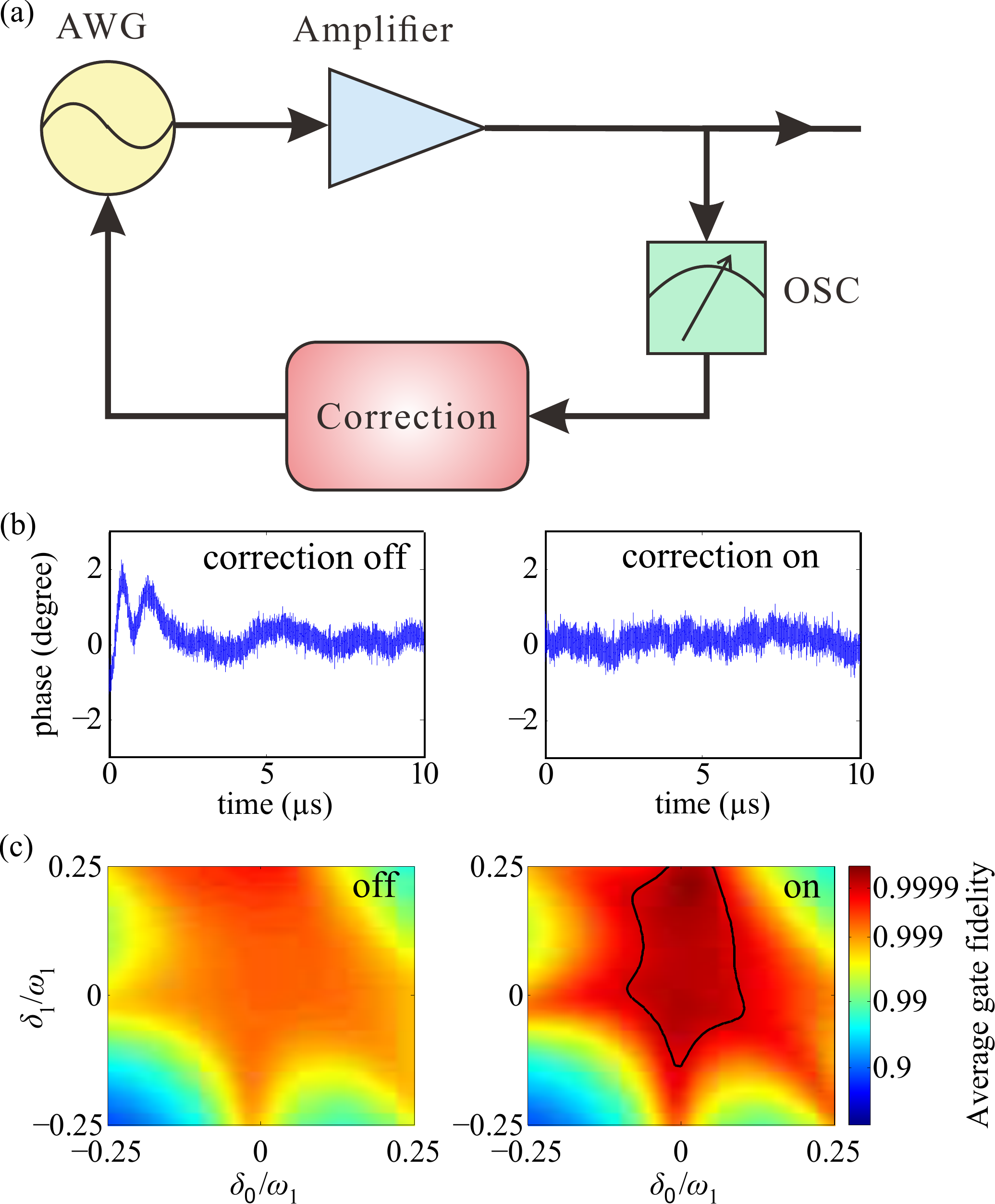}
\renewcommand{\figurename}{Supplementary Figure}
\caption{\label{MWCal} \textbf{Correction of microwave pulse distortions.} (a) Schematic correction diagram showing the main instruments generating, processing, and sampling the microwave pulses. (b) Extracted phase distortions without (left penal) and with (right penal) the correction. (c) Calculated average gate fidelity of a BB1inC $\pi/2$ gate with sampled waveforms without (left penal) and with (right penal) the correction. The region of fidelity larger than 0.9999 is labeled.}
\end{figure*}

\begin{figure*}[h]
\centering
\includegraphics[width=0.8\textwidth]{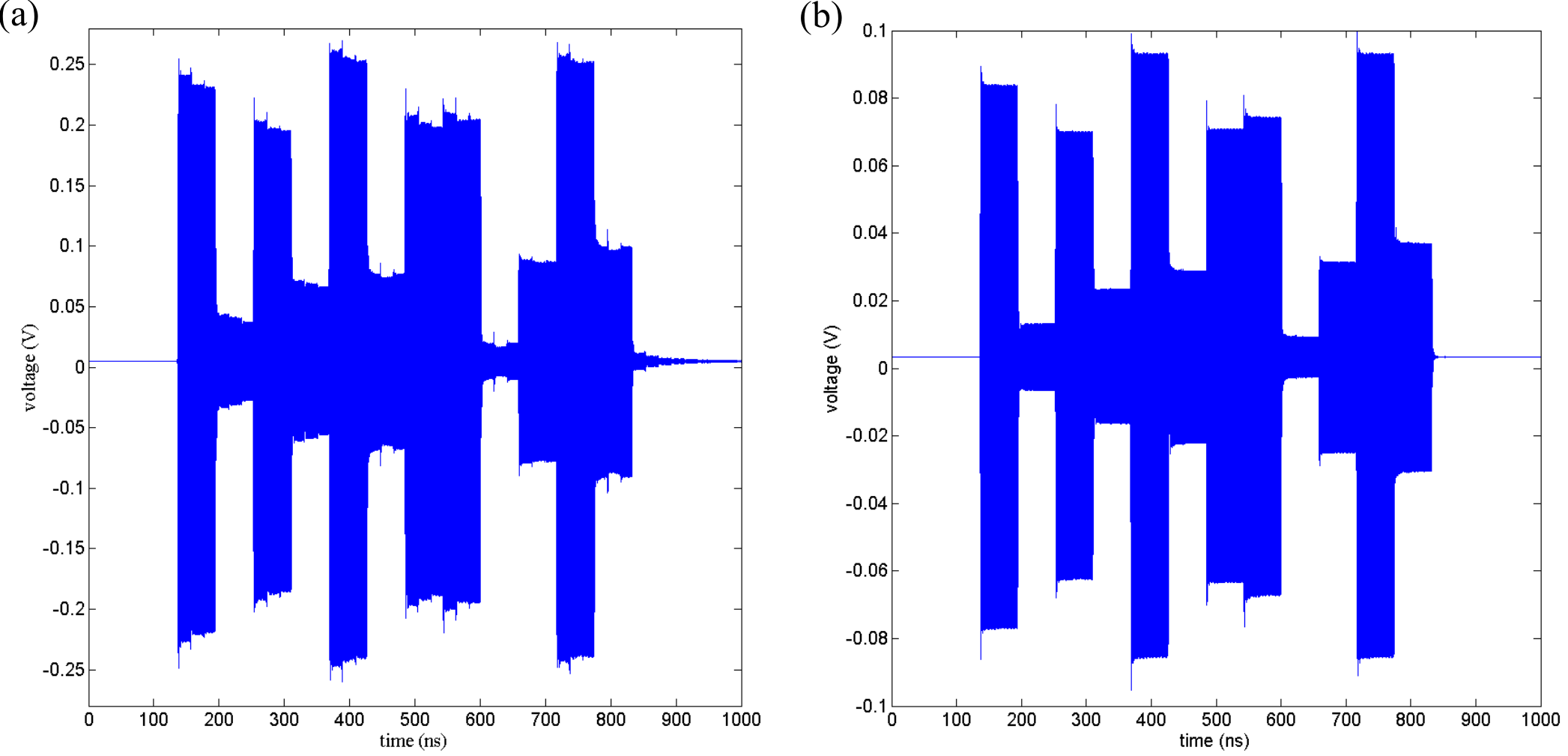}
\renewcommand{\figurename}{Supplementary Figure}
    \caption{\textbf{Correction of pulse distortions caused by leakage and reflection.} (a) Waveform of a GRAPE pulse without improvement. The distortions are mainly caused by leakage and reflection of the diplexer. (b) Improved waveform of the pulse sequence. The distortions are corrected by inserting a 10 dB attenuator between microwave components.
 }
    \label{ReflectCorrect}
\end{figure*}

\begin{figure*}[h]
\centering
\includegraphics[width=0.8\textwidth]{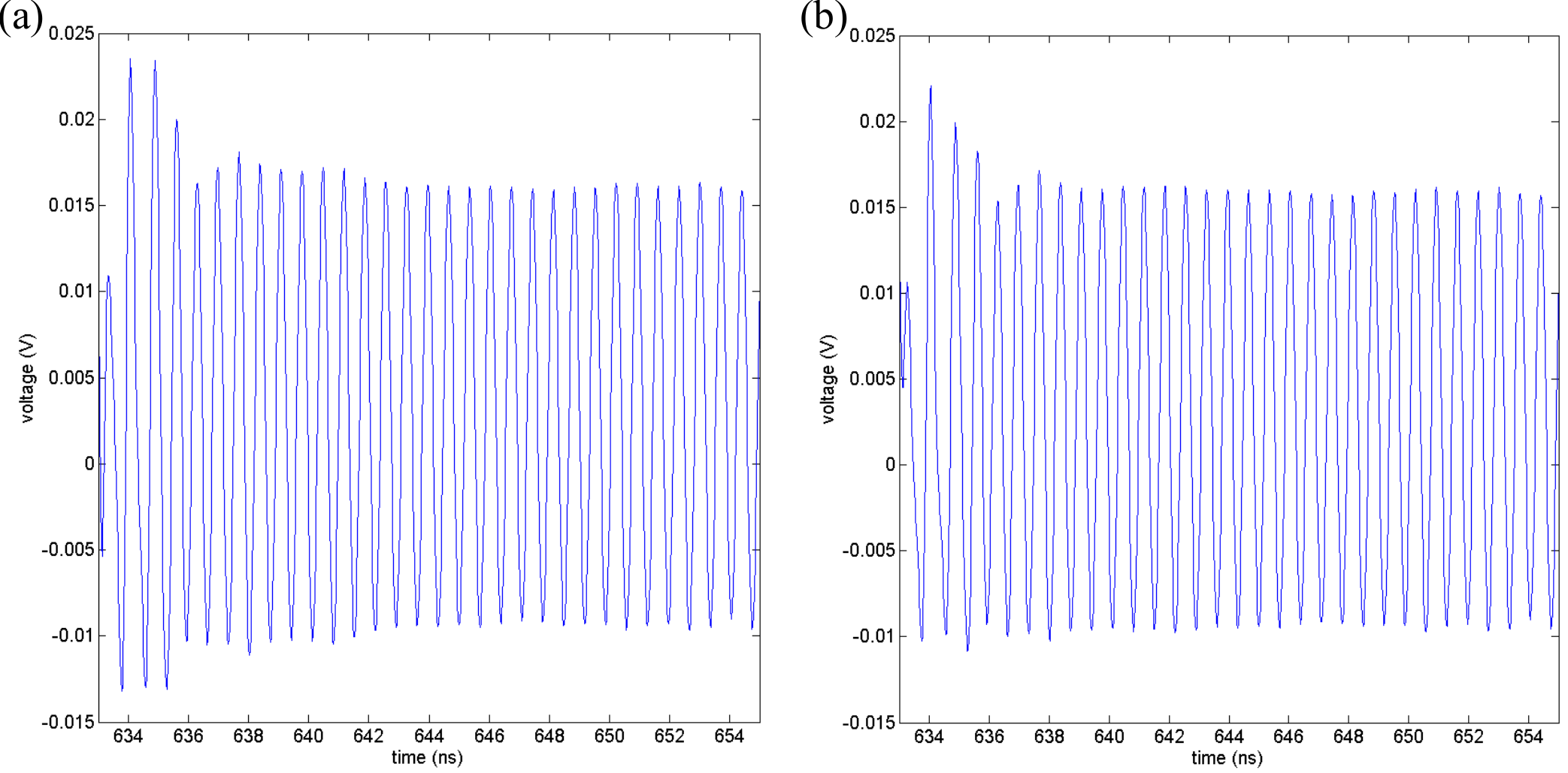}
\renewcommand{\figurename}{Supplementary Figure}
    \caption{\textbf{Correction of the amplitude distortions.} (a) Waveform without correction. (b) Waveform with correction.
 }
    \label{AmpliCalib}
\end{figure*}

\begin{figure*}[h]
\centering
\includegraphics[width=0.8\textwidth]{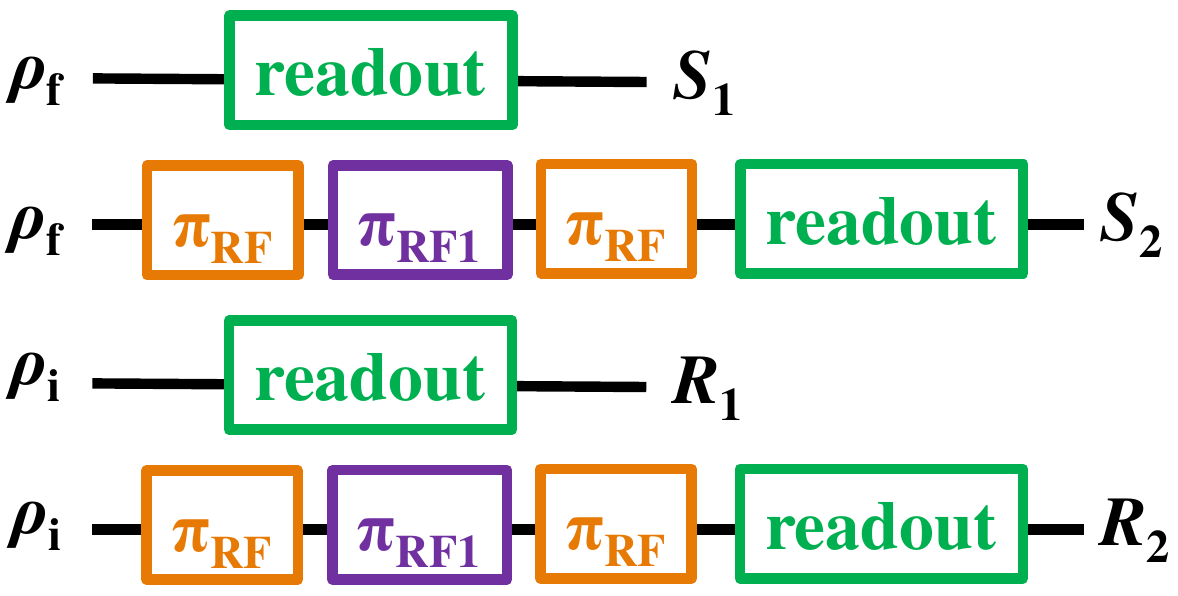}
\renewcommand{\figurename}{Supplementary Figure}
    \caption{\textbf{Schematic normalization sequences in the two-qubit experiment.} Here $\rho_\textrm{i}$ denotes the initialized state with laser, $\rho_\textrm{f}$ denotes the final state after applying control sequence to $\rho_\textrm{i}$. $\pi_{\textrm{RF}}$ ($\pi_{\textrm{RF1}}$) is a radio-frequency $\pi$ pulse driving the nuclear spin transition between states $|m_S=0, m_I=1\rangle$ and $|m_S=0, m_I=0\rangle$ (transition between states $|m_S=0, m_I=0\rangle$ and $|m_S=0, m_I=-1\rangle$). The measured photoluminescence intensity after each sequence is denoted by $S_1$, $S_2$, $R_1$, or $R_2$.
 }
    \label{TwoQubitNorm}
\end{figure*}

\begin{figure*}[h]
\centering
\includegraphics[width=0.8\textwidth]{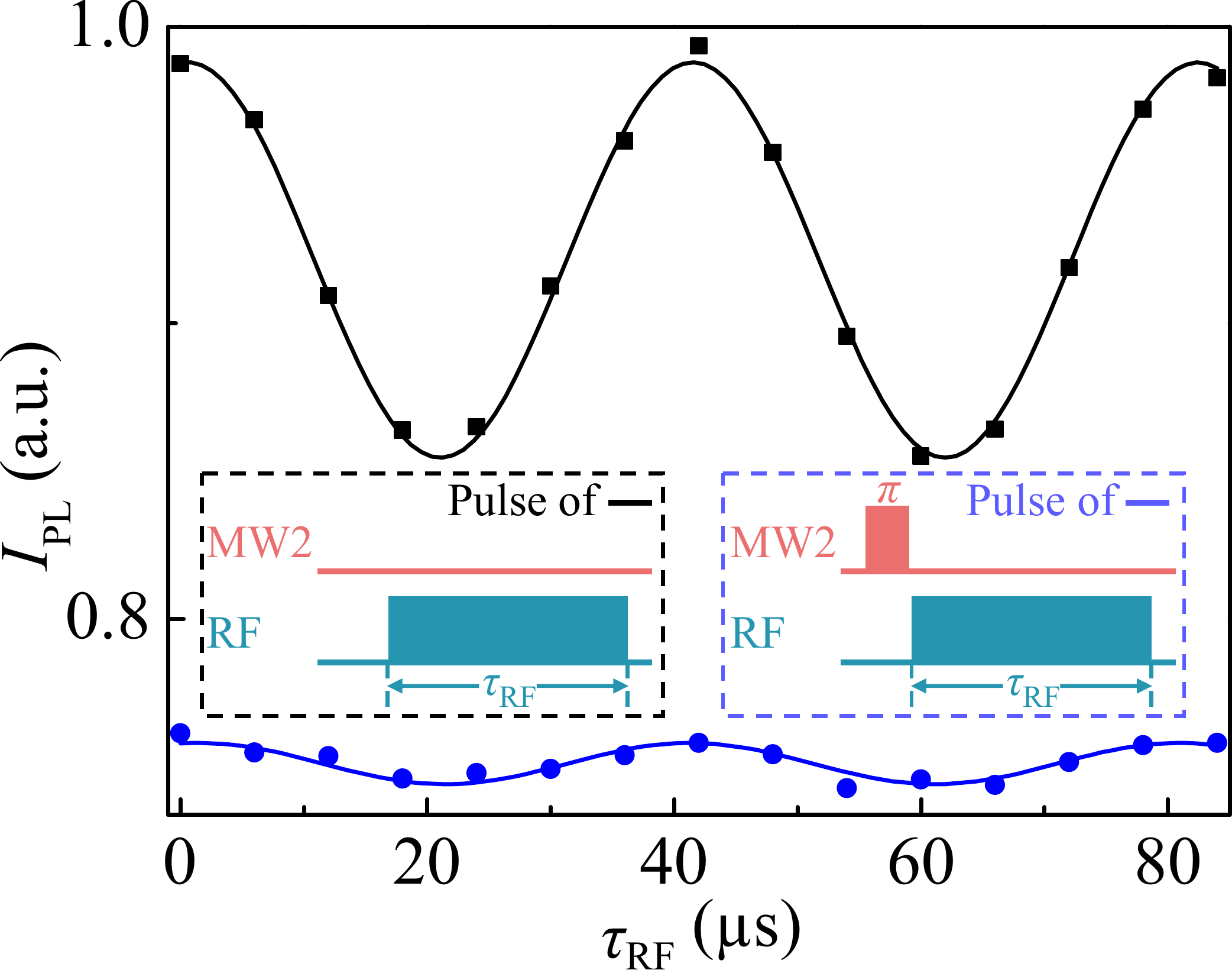}
\renewcommand{\figurename}{Supplementary Figure}
    \caption{\textbf{Experimental measurement of the polarization of the NV electron spin.} The polarization $\alpha$ is extracted from the ratio of the two nutation amplitudes to be 0.83(2).
 }
    \label{PolarNV}
\end{figure*}

\begin{figure*}[h]
\centering
\includegraphics[width=0.8\textwidth]{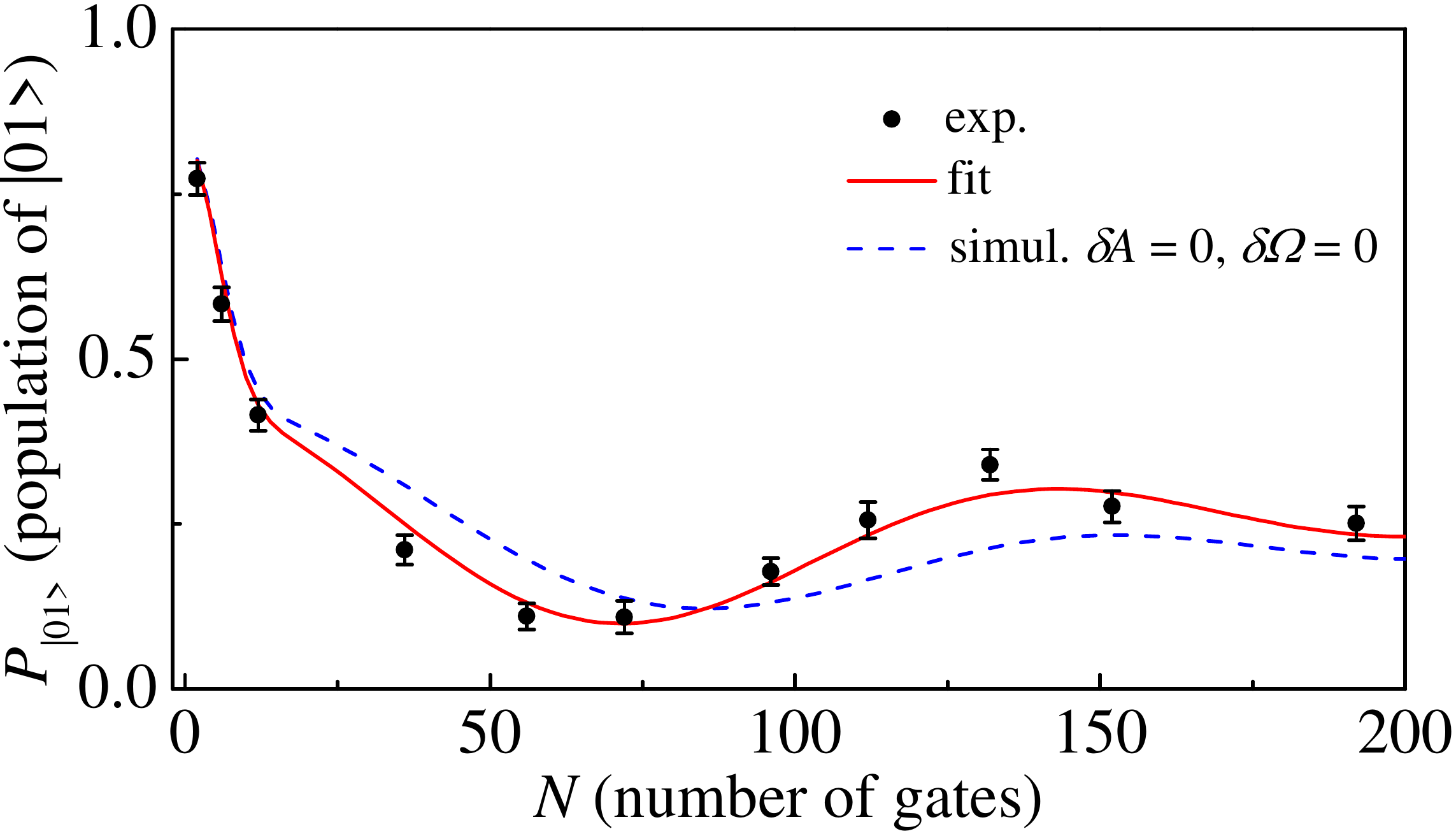}
\renewcommand{\figurename}{Supplementary Figure}
    \caption{\textbf{Extraction of the practical Hamiltonian.} The experimentally measured $P_{|01\rangle}$ is shown as the black circles. The practical Hamiltonian is extracted by fitting the experimental data. The fitted $P_{|01\rangle}$ is shown as the red solid line. Best-fit values of $\delta A$ and $\delta\Omega$ are $\delta A=0.008(1)~$MHz and $\delta\Omega=0.068(1)~$MHz. For comparison, the blue dashed line shows the simulated result with $\delta A=0$ and $\delta\Omega=0$.
 }
    \label{P01compHamil}
\end{figure*}

\begin{figure*}[h]
\centering
\includegraphics[width=0.8\textwidth]{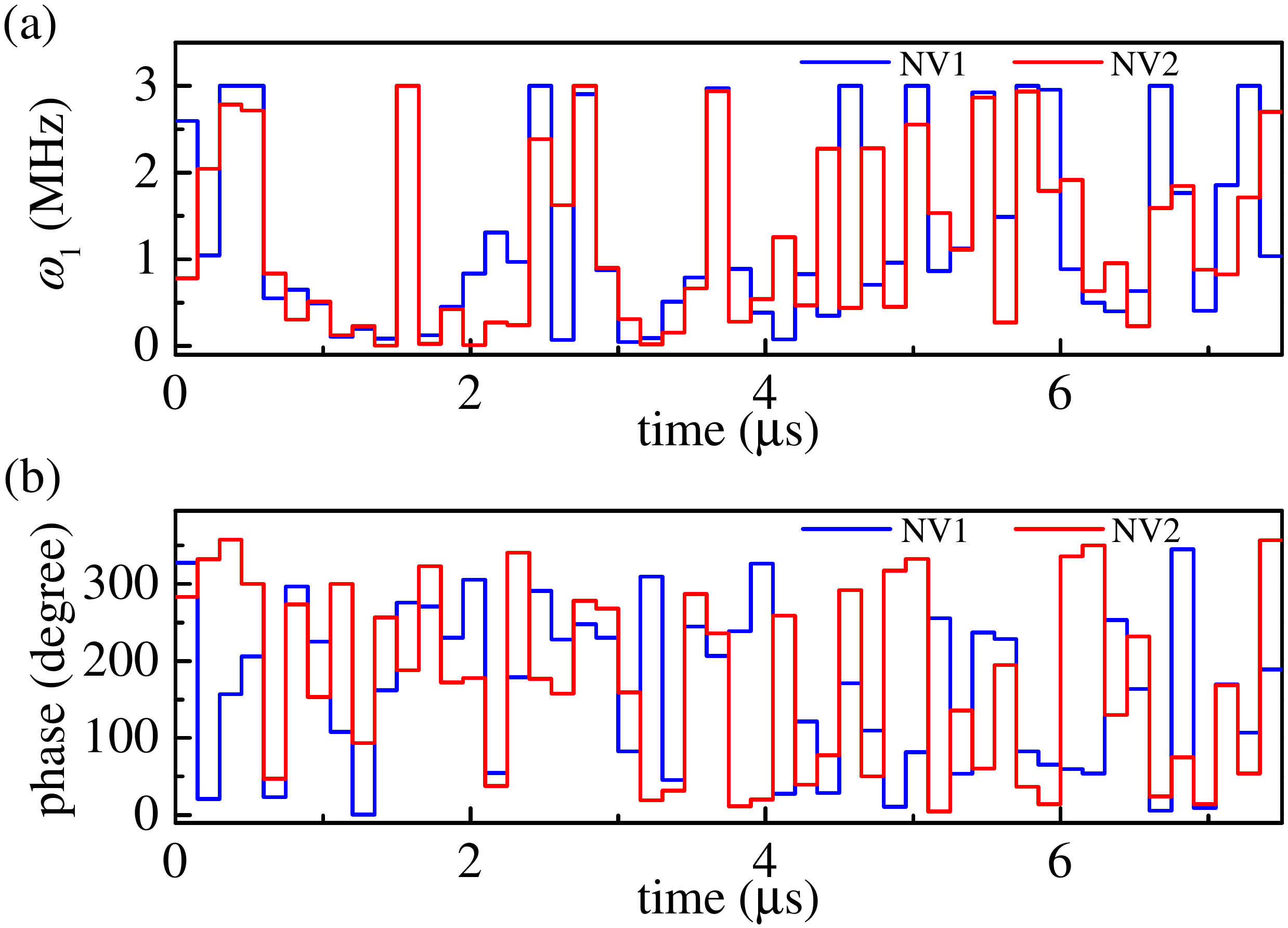}
\renewcommand{\figurename}{Supplementary Figure}
    \caption{\textbf{Pulse sequence for the CNOT gate in NV-NV coupled system.} (a) Amplitudes of the control microwave fields. (b) Phases of the control microwave fields.
 }
    \label{NVNVCNOTseq}
\end{figure*}

\clearpage

\section{Supplementary Tables}
\begin{table}[h]
\centering  
\renewcommand{\tablename}{Supplementary Table}
\caption{\textbf{Summarization of the results of single-qubit randomized benchmarking.} Average gate fidelity ($F_{\textrm{a}}$) and error per gate ($\varepsilon_{\textrm{g}}$) is shown for naive, five-piece SUPCODE, BB1, and BB1inC pulses. }
\textrm{\\}
\begin{tabular}{c|cc}  
\hline\hline
Pulse sequence &$F_\textrm{a}$ &$\varepsilon_\textrm{g}$\\ \hline  
naive &0.99968(6) &3.2(6)$\times$10$^{-4}$\\         
five-piece SUPCODE &0.99916(8) &8.4(8)$\times$10$^{-4}$\\        
BB1 &~~0.999945(6) &~5.5(6)$\times$10$^{-5}$\\        
BB1inC &~~0.999952(6) &4.8(6)$\times$10$^{-5}$\\ \hline\hline
\end{tabular}
\label{RBresults}
\end{table}

\section{Supplementary Notes}

\subsection{Supplementary Note 1}
\textbf{Hamiltonian of the NV system.} The NV center includes a
substitutional nitrogen atom and a vacancy in the nearest-neighbor
lattice position. In our experiment,  a static magnetic field, $B_0=513~$ G, is applied along the NV symmetry axis ([1 1 1] crystal axis).
The Hamiltonian of the NV center can be written as
\begin{equation}
\label{NVHamilt}
H_{\textrm{NV}} =  2\pi(DS_z^{2}+\omega_S S_z
 +PI_z^{2} - \omega_I I_z) + H_{\textrm{hf}},
\end{equation}
where $\omega_S = -\gamma_\textrm{e} B_0/2\pi$ ($\omega_I = \gamma_\textrm{N} B_0/2\pi$) is the Zeeman splitting of the electron ($^{14}$N nuclear) spin, $\gamma_\textrm{e}$ ($\gamma_\textrm{N}$) is the electronic ($^{14}$N nuclear) gyromagnetic ratio, $S_z$ and $I_z$ are the electron and nitrogen nuclear spin operators, respectively. The zero field splitting $D = 2870~$MHz and the nuclear quadrupolar splitting $P = -4.95$~MHz. The hyperfine interaction between the NV electron spin and the $^{14}$N nuclear spin is
\begin{equation}
\label{Hhf}
H_{\textrm{hf}} =  2\pi[A_\perp(S_xI_x+S_yI_y)+AS_zI_z],
\end{equation}
The strength of the hyperfine interaction is about 2~MHz. Because of the strong zero field splitting and Zeeman splitting terms of the electron spin, the effect of the interaction term $S_xI_x+S_yI_y$ is strongly suppressed and can be neglected. $A=-2.16$~MHz is determined via CW ESR experiment.
In the secular approximation, the Hamiltonian is
\begin{equation}
\label{NVHamiltpara}
H_{\textrm{NV}} =  2\pi(DS_z^{2}+\omega_S S_z
+ A S_z I_z +PI_z^{2}
   - \omega_I I_z),
\end{equation}
The electron (nuclear) spin states $|m_S = 0\rangle$ and $|m_S = -1\rangle$ ($|m_I = 0\rangle$ and $|m_I = +1\rangle$ ) are encoded as the electron (nuclear) spin qubit.

 Microwave (MW) and radio-frequency (RF) pulses are used to manipulate the two-qubit system. The frequency of MW and RF pulses are $f_{\textrm{MW}}$ and $f_{\textrm{RF}}$
, respectively.
When MW pulses are applied, the total Hamiltonian becomes
\begin{equation}
\label{Htot}
H =  H_{\textrm{NV}}+H_\textrm{C},
\end{equation}
with
\begin{equation}
\label{HC}
H_\textrm{C} =  2\pi\sqrt{2}\omega_1\cos(2\pi f_{\textrm{MW}}t+\phi)S_x,
\end{equation}
where $\phi$ is the phase of the MW pulse, $\omega_1$ is the amplitude of the MW pulse.

The Hamiltonian can be transformed into the rotating frame as
\begin{equation}
\label{Hrot}
H_{\textrm{rot}} =  U_{\textrm{trans}}HU_{\textrm{trans}}^\dag-iU_{\textrm{trans}}\frac{dU_{\textrm{trans}}^\dag}{dt},
\end{equation}
with
\begin{equation}
\label{Utrans}
U_{\textrm{trans}} =  e^{i2\pi f_{\textrm{MW}}tS_z^2}e^{-i2\pi f_{\textrm{RF}}tI_z^2}.
\end{equation}
With rotating-wave approximation, the Hamiltonian in the rotating frame can be simplified as
\begin{equation}
\label{Hrotexp}
\begin{aligned}
&H_{\textrm{rot}} =  2\pi(\omega_S(S_z+S_z^2)-\omega_I(I_z-I_z^2)+A(S_z^2+S_zI_z)) \\
&+H_{\textrm{C, rot}},
\end{aligned}
\end{equation}
where
\begin{equation}
\label{HCrot}
H_{\textrm{C, rot}} =  2\pi[\delta\Omega S_z^2 +  \omega_1/\sqrt{2}(\cos\phi H_x+\sin\phi H_y)],
\end{equation}
with
\begin{equation}
\label{Hx}
H_x =  \frac{1}{\sqrt{2}}\left(
  \begin{array}{ccc}
    0 & 1 & 0\\
    1 & 0 & 1\\
    0 & 1 & 0\\
  \end{array}
\right)\otimes \mathbb{I} ,
\end{equation}
\begin{equation}
\label{Hy}
H_y =  \frac{1}{\sqrt{2}}\left(
  \begin{array}{ccc}
    0 & -i & 0\\
    i & 0 & i\\
    0 & -i & 0\\
  \end{array}
\right)\otimes \mathbb{I}  ,
\end{equation}
 , $\delta\Omega = D - \omega_S -A - f_{\textrm{MW}}$, $f_{\textrm{RF}}=-P+\omega_I$, and $\mathbb{I}$ representing $3\times3$ identity matrix.

\subsection{Supplementary Note 2}
\textbf{Calculation of average gate fidelity.} The average gate fidelity between a quantum operation $\xi$ and a target unitary quantum gate $U$ is defined as
\begin{equation}
\label{Fave}
F_{\textrm{a}}(\xi, U) =  \int d\psi \langle\psi|U^\dag \xi(|\psi\rangle\langle\psi|)U|\psi\rangle,
\end{equation}
where the integral is over the uniform measure $d\psi$ on state space, normalized so $\int d\psi=1$ \cite{PLA303_249}.

In the single-qubit case, the average gate fidelity can be derived to be \cite{PLA294_258}
\begin{equation}
\label{Favesingle}
F_{\textrm{a}}^{(1)}(\xi, U) =  \frac{1}{2}+\frac{1}{12}\sum_{j=x, y, z}tr(U\sigma_jU^\dag\xi(\sigma_j)),
\end{equation}
where $\sigma_x$, $\sigma_y$, and $\sigma_z$ are Pauli matrices.

Quantum optimal control method \cite{JMR172_296} is used to design the pulse sequence of CNOT gate. To calculate the average gate fidelity of this CNOT gate, Eqn. \ref{Fave} is generalized so that the integral is on the two-qubit space \cite{PLA367_47}. The nuclear spin is much less sensitive to the external magnetic noise than the electron spin and the GRAPE pulse sequence contains only microwave pulses, so the decoherence during the operation mainly comes from the static distributions of $\delta_0$ and $\delta_1$ for the electron spin qubit. Then the operation can be expressed as
\begin{equation}
\label{varepsilon}
\begin{aligned}
&\xi(|\psi\rangle\langle\psi|) =  \\
&\int d\delta_0\int d\delta_1P_0(\delta_0)P_1(\delta_1)U_{\textrm{seq}}(\delta_0, \delta_1)|\psi\rangle\langle\psi| U_{\textrm{seq}}^\dag(\delta_0, \delta_1),
\end{aligned}
\end{equation}
where $U_{\textrm{seq}}(\delta_0, \delta_1)$ is the calculated two-qubit evolution according to the pulse sequence, with the errors $\delta_0$ and $\delta_1$ considered in Hamiltonian. Substituting Eqn. \ref{varepsilon} into Eqn. \ref{Fave} yields the average gate fidelity between the operation $\xi$ and the target CNOT gate $U_{\textrm{CNOT}}$,
\begin{equation}
\label{FaveCNOT}
\begin{aligned}
&F_{\textrm{a}}(\xi, U_{\textrm{CNOT}}) = \\
&\frac{1}{d(d+1)} \int d\delta_0\int d\delta_1P_0(\delta_0)P_1(\delta_1)(tr(MM^\dag)+|tr(M)|^2),
\end{aligned}
\end{equation}
with $d=4$  and
\begin{equation}
\label{M}
M =  U_{\textrm{CNOT}}^\dag U_{\textrm{seq}}(\delta_0, \delta_1).
\end{equation}
It can be easily obtained from Eqn. \ref{FaveCNOT} that the fidelity of operation $\xi$ without the effect of the noise ($\delta_0$ and $\delta_1$) can be written as
\begin{equation}
\label{Fseq}
F_{\textrm{seq}} =  \frac{1}{d(d+1)}[tr(MM^\dag)+|tr(M)|^2],
\end{equation}
where the values of $\delta_0$ and $\delta_1$ are zero.

\subsection{Supplementary Note 3}
\textbf{High fidelity single-qubit quantum gates.} Considering a single-qubit gate corresponding to a rotation of angle $\theta$ around the $x$ axis on the Bloch sphere, such a gate can be realized by evolution under the effective Hamiltonian $H_{\textrm{ideal}} = 2\pi \omega_1 \mathbf{n}\cdot \mathbf{S}$,
where $\mathbf{S}=(S_x,S_y,S_z)$ is the spin vector operator of the qubit, $\mathbf{n}$
is a three-dimensional vector, and the strength $\omega_1$ is a real parameter.
The average gate fidelity is limited by interaction of the qubit with environment and fluctuation of the control field.
We consider the model where the Hamiltonian for rotation about the $x$ axis under practical conditions is described as $H_{\textrm{prac}}=2\pi\delta_0S_z+2\pi(\omega_1+\delta_1)[\cos \delta\phi S_x + \sin \delta\phi S_y]$.
 The error $\delta_0$ in the Hamiltonian is due to the interaction of the qubit with environment,  the error $\delta_1$ is due to fluctuation of the control field strength and phase error $\delta\phi$ is caused by the imperfect microwave pulse generation.
 Phase error can be efficiently eliminated by pulse fixing technique (detailed in Section \ref{DistortCorrect}) and we take it as of zero value in this section.
 We consider the case where both $\delta_0$ and $\delta_1$ vary in a timescale much longer than that of quantum gates. In this case $\delta_0$ and $\delta_1$ are taken as quasi-static random constants.
\label{SingleQubit}

Supplementary Figure \ref{Pulsedata}a shows the performance of the gate by simply applying a naive rectangular pulse. Here the gate $(\pi/2)_0$ (we denote the rotation of an angle $\theta$ around the axis in the equatorial plane with azimuth $\phi$ as $(\theta)_{\phi}$) is taken as an example. The average gate fidelity of $(\pi/2)_0$ is calculated with respect to different values of $\delta_0$ and $\delta_1$. The naive pulse is vary sensitive to the errors $\delta_0$ and $\delta_1$, with leading orders of both errors preserved in the evolution operator (corresponding to second orders in the average gate fidelity). This corresponds with the small region of high average gate fidelity shown in the lower panel of Supplementary Figure \ref{Pulsedata}a.

Supplementary Figure \ref{Pulsedata}b shows a type of dynamically corrected gate, five-piece SUPCODE \cite{NatCommun3_997}. The pulse sequence is depicted as $\tau_1-(\theta/2)_0-\tau_3-(\theta/2)_0-\tau_1$ for $\theta\in(2\pi, 3\pi)$. Here $\tau_1=\csc\theta(1-2\cos\frac{\theta}{2}+\cos\theta+\sqrt{4-8\cos\frac{\theta}{2}+4\cos\theta+\theta \sin\theta})$ and $\tau_3=-2(\tau_1\cos\frac{\theta}{2}+\sin\frac{\theta}{2})$ are durations when control field is off. Under the five-piece SUPCODE pulse, up to second order of $\delta_0$ can be canceled (corresponding to sixth order preserved in the average gate fidelity). In the lower panel of Supplementary Figure \ref{Pulsedata}b, the average gate fidelity of $(2.5\pi)_0$ (equivalent to $(\pi/2)_0$ in the single-qubit case) is shown as an example. The region of high average gate fidelity is largely extended in the axis of $\delta_0$, compared with that by the naive pulae.

In Supplementary Figure \ref{Pulsedata}c a type of composite pulse, BB1 \cite{JMRA109_221}, is shown. The pulse sequence is $(\theta/2)_0-(\pi)_{\phi}-(2\pi)_{3\phi}-(\pi)_{\phi}-(\theta/2)_0$, with $\phi=\arccos(-\theta/4\pi)$. Under the BB1 pulse, up to second order of $\delta_1$ is canceled in the evolution operator (corresponding to sixth order preserved in the average gate fidelity). The average gate fidelity of $(\pi/2)_0$ with the BB1 pulse is shown in the lower panel of Supplementary Figure \ref{Pulsedata}c. It exhibits a larger region of high fidelity in the axis of $\delta_1$. Thus suppressing the $\delta_1$ error by applying the BB1 pulse, in combination with a proper selection of control field strength $\omega_1$ to suppress the $\delta_0$ error, can contribute to realization of a high fidelity (e.g. 0.9999).

Supplementary Figure \ref{Pulsedata}d shows a pulse sequence suppressing both the $\delta_0$ and $\delta_1$ errors simultaneously. The sequence is designed by incorporating the BB1 pulse within the CORPSE pulse, and is named BB1inC here for short. There are similar pulse sequences to suppress the $\delta_0$ and $\delta_1$ error simultaneously \cite{JPSJ82_014004}. The BB1inC sequence is depicted as $(\theta/2)_0-(\pi)_{\phi}-(2\pi)_{3\phi}-(\pi)_{\phi}-(\theta_3)_0-(\theta_2)_{\pi}-(\theta_1)_0$, where $\phi=\arccos(-\theta/4\pi)$, $\theta_1=\theta/2-\arcsin(\sin(\theta/2)/2)$, $\theta_2=2\pi-2\arcsin(\sin(\theta/2)/2)$, and $\theta_3=2\pi-\arcsin(\sin(\theta/2)/2)$. Leading orders of both the $\delta_0$ and $\delta_1$ errors are canceled in the evolution operator. The lower panel of Supplementary Figure \ref{Pulsedata}d shows the average gate fidelity of $(\pi/2)_0$ with the BB1inC pulse. The region of high fidelity is much larger, extended in both axes of $\delta_0$ and $\delta_1$.

Recently the robustness of composite pulse sequences against time-dependent noise is analyzed \cite{PRA90_012316}. It is shown that composite pulses may also be successfully employed in the presence of time-dependent noise. The robustness against static as well as time-dependent noise enables composite pulse an effective method to improve single-qubit gate fidelity.

\subsection{Supplementary Note 4}
\textbf{Quantum optimal control method for designing the CNOT gate.} GRAPE is a type of quantum optimal control method. It can be utilized to design control sequence to realize a target gate with high fidelity.
The control sequence contains $N$ piece of pulses, with the amplitude and phase of each piece being different.
The total Hamiltonian of the $k$th pulse in the rotating frame is (see Eqn. \ref{Hrotexp} and Eqn. \ref{HCrot})
\begin{equation}
\label{Hrotk}
\begin{aligned}
&H_{\textrm{rot,} k} =  2\pi(\omega_S(S_z+S_z^2)-\omega_I(I_z-I_z^2)+A(S_z^2+S_zI_z))\\
&+H_{\textrm{C, rot,} k},
\end{aligned}
\end{equation}
with
\begin{equation}
\label{HCrotk}
H_{\textrm{C, rot,} k} =  2\pi\omega_{1, k}/\sqrt{2}(\cos\phi_k H_x+\sin\phi_k H_y),
\end{equation}
where $\phi_k$ is the phase of the $k$th pulse, $\omega_{1, k}$ is the amplitude of the $k$th microwave pulse and $\delta\Omega = 0$.

The evolution operator under the $k$th pulse is written as
\begin{equation}
\label{Uk}
U_{k} =  e^{-iH_{\textrm{rot,} k}t_k},
\end{equation}
where $t_k$ is the duration of the $k$th pulse. The total evolution under the entire sequence is
\begin{equation}
\label{Useq}
U_{\textrm{tot}} =  \prod_{k=N}^1U_k.
\end{equation}
The time duration of each pulse, $t_k$, is set to be equal value $\tau$. The two-qubit evolution operator can be described as
\begin{equation}
\label{Useq}
U_{\textrm{seq}}(\{\omega_{1, k}, \phi_k\}) =  \mathcal{P}\prod_{k=N}^1e^{-i H_{\textrm{rot,} k}(\{\omega_{1, k}, \phi_k\})\tau}\mathcal{P},
\end{equation}
where $\mathcal{P}$ is the projection operator on the two-qubit subspace.

\label{optmcont}
The target gate is the CNOT gate,
\begin{equation}
\label{UCNOT}
U_{\textrm{CNOT}} = \left(
  \begin{array}{cccc}
    0 & 0 & 1 & 0\\
    0 & 1 & 0 & 0\\
    1 & 0 & 0 & 0\\
    0 & 0 & 0 & 1\\
  \end{array}
\right).
\end{equation}
The performance function of the GRAPE algorithm is the fidelity $F_{\textrm{seq}}$, which is a function of $\{\omega_{1, k}, \phi_k\}$. The values of $\{\omega_{1, k}, \phi_k\}$ are initialized with random numbers within the experimental conditions. The performance function are maximized by iteration. Within each iteration, the value of $F_{\textrm{seq}}(\{\omega_{1, k}, \phi_k\})$ as well as its derivative to $\omega_{1, k}$ and $\phi_k$ is calculated, then the value of $\omega_{1, k}$ ($\phi_k$) is replaced by the result of its previous value plus the derivative multiplied by a proper coefficient.

This method works well for designing a sequence with high fidelity, if errors due to qubit-environment interaction and control field fluctuation are not taken into account.
However, our aim is to realize quantum gates, which are not only of high fidelity, but also being robust to the errors.
A method has been presented \cite{JMR172_296} to design pulse sequence which is robust against the inhomogeneity of $\omega_1$.
Herein, we generalize this method to design pulse sequence, which is not only robust against the inhomogeneity  $\delta_1$, but also is insensitive to the dephasing noise $\delta_0$.
The performance function of the modified GRAPE is defined as $F_\textrm{a}(\varepsilon, U_{\textrm{CNOT}})$  (see Eqn. \ref{FaveCNOT}).
The new performance function is maximized by iteration of the GRAPE algorithm. In practical implementation of the modified algorithm, the integral in Eqn. \ref{FaveCNOT} is replaced by sum of discrete points. We find that three points of $\delta_0$ ($\delta_{1}$) are enough.

Supplementary Figure \ref{GRAPEseqFave}a shows the amplitude and phase of the designed pulse sequence. The sequence consists of twelve pieces of pulse. The duration of each pulse is $58~$ns. Without considering the errors $\delta_0$ and $\delta_1$, the sequence produces a two-qubit operation $U_{\textrm{cal}}$,
\begin{equation}
\label{Ucalsub}
\begin{aligned}
&U_{\textrm{cal}} = e^{1.5974i}\left(
  \begin{array}{cccc}
    -0.0060 & 0 & 1 & 0\\
    0 & 0.9996 & 0 & 0.0154\\
    0.9999 & 0 & 0.0059 & 0\\
    0 & -0.0146 & 0 & 0.9991\\
  \end{array}
\right) \\
~\\
& +ie^{1.5974i}\left(
  \begin{array}{cccc}
    0.0032 & 0 & -0.0013 & 0\\
    0 & 0 & 0 & -0.0249\\
    0.0119 & 0 & 0.0032 & 0\\
    0 & -0.0253 & 0 & 0.0316\\
  \end{array}
\right),
\end{aligned}
\end{equation}
The fidelity of $U_{\textrm{cal}}$ is $0.9995$. Supplementary Figure \ref{GRAPEseqFave}b shows the robustness of the sequence against the errors $\delta_0$ and $\delta_1$. When the experimental distributions $P_0(\delta_0)$ and $P_1(\delta_1)$ (which are determined from the experiments) are considered, the sequence provides an average gate fidelity of $0.9927$.

\subsection{Supplementary Note 5}
\textbf{Alignment of the magnetic field.} We used the fluorescence dependence on the misalignment angle to align the magnetic field. According to the literature \cite{Awschalom2005NatPhys}, the fluorescence of NV center is sensitive to misalignment angle of the NV axis from a magnetic field $\textbf{B}_0$ when the magnitude is approximately $513~$G. The difference in fluorescence counts is still noticeable even when the misalignment angle is only 0.5$^\circ$. In our experiment, the fluorescence count was the same (within counting errors) for $B_0 \approx 0~G ~\text{and}~ B_0 \approx 513~$G. So we estimate the misalignment angle to be within $0.5^\circ$.

\textbf{Creation of a solid immersion lens.} All measurements in our experiment are based on detection of the NV photoluminescence. Much of the photoluminescence is lost at the diamond surface due to internal reflection. The problem can be overcome by creating a solid immersion lens (SIL) \cite{Nat506_204}. We created a SIL in the diamond around an NV center (Supplementary Figure \ref{SIL}). The SIL increases the PL rate to about $400~\textrm{kcounts s}^{-1}$.

\textbf{Ultra-broadband coplanar waveguide.} In the experiment, manipulation of qubits is achieved by microwave (MW) and radio-frequency (RF) pulses, which are applied through a coplanar waveguide (CPW). The ultra-broadband CPW is designed and fabricated. Supplementary Figure \ref{CPWpara} shows scattering parameters of the CPW. Up to 15 GHz, the S21 parameter is larger than -3 dB, and the S11 parameter is about or less than -10 dB. Such wide bandwith ensures that there is almost no extra distortion of microwave / RF pulses with this CPW.

\subsection{Supplementary Note 6}
\label{DistortCorrect}
The imperfect devices generate microwave pulses with non-ideal amplitudes and phases.
The imperfection of microwave pulses sent to the NV centers are carefully corrected with pulse fixing technique \cite{PRL109_070504}.
Supplementary Figure \ref{MWCal}a  shows the main instruments for the microwave pulse generation. The pulses are generated by an arbitrary waveform generator (M8190A, Keysight ), and amplified with a power amplifier (ZHL-30W-252-S+, Mini Circuits). The imperfections in the instruments cause distortions of the microwave pulses, which may dramatically decrease fidelity of quantum gates. An oscilloscope (DSO-X 92004Q) is used to sample the microwave pulses. The pulse distortions are then corrected by predistorting the pulses in the right way that the predistortions cancel with the distortions by the imperfections of instruments.

\textbf{Microwave phase correction.}
Supplementary Figure \ref{MWCal}b shows the distortion of the microwave pulse phase with / without the correction. It is clear that there is no significant distortion of the microwave pulse phase with the correction, as shown in the right panel of Supplementary Figure \ref{MWCal}b. In Supplementary Figure \ref{MWCal}c, we compare  the average gate fidelity theoretically with the microwave pulse with and without the phase correction. Because of the pulse distortions, the BB1inC $\pi/2$ gate becomes less error-resilient, without any region of fidelity higher than 0.9999 (see the left panel of Supplementary Figure \ref{MWCal}c ). The result in  the right panel of Supplementary Figure \ref{MWCal}c shows a region with fidelity higher than $0.9999$ with the phase correction.

\textbf{Microwave amplitude correction.}
In the two-qubit experiment, microwave and radio-frequency pulses are combined with a diplexer (Marki DPX-1). We find that the leakage and reflection of the diplexer ports cause extra distortions of the microwave pulses. Supplementary Figure \ref{ReflectCorrect}a shows that there are distortion of the microwave amplitude. After inserting a 10~dB attenuator between the  microwave components to suppress the leakage and reflection, the waveform of the pulse sequence is improved to be close to the ideal case, as shown in Supplementary Figure \ref{ReflectCorrect}b.

The distortions of amplitude shown in Supplementary Figure \ref{ReflectCorrect}b can be further corrected by pulse fixing technique. Similar to that shown in Supplementary Figure \ref{MWCal}a, the distortions are recorded by an oscilloscope, and then fed back to the arbitrary waveform generator so that the distortions are minimized. Supplementary Figure \ref{AmpliCalib} shows the comparison of the pulse waveforms without and with amplitude correction.

\subsection{Supplementary Note 7}
\textbf{Normalization of the experimental data.} In the single-qubit experiment, the normalization is carried out by performing a nutation experiment \cite{PRL112_050503}. The normalized data corresponds to the population of $|0\rangle$ for the final state.

In the two-qubit experiment, the population of $|m_S=0, m_I=1\rangle$ ($P_{|01\rangle}$) for the final state is obtained by normalization.
According to the Ref. \onlinecite{Nat484_82}, each occupied energy level contributes to the measured photoluminescence intensity ($I_{\textrm{PL}}$) with a different PL rate and these different PL rates  are measured and used to determine the population of the levels with several sequences. Herein we introduce an alternative method for normalization. The pulse sequences for the normalization are shown in Supplementary Figure \ref{TwoQubitNorm}. The measured $I_{\textrm{PL}}$  is
\begin{equation}
\label{IPL}
I_{\textrm{PL}} =  \sum_{|k\rangle}\beta_{|k\rangle}P_{\rho, |k\rangle},
\end{equation}
where $|k\rangle$ denotes the nine energy levels of the NV center ($|m_S, m_I\rangle$ with $m_S=0, \pm1$ and $m_I=0, \pm1$), $P_{\rho, |k\rangle}$ is the population of $|k\rangle$ for the state $\rho$, $\beta_{k}$ is the photoluminescence intensity if the state is $|k\rangle$.

In Supplementary Figure \ref{TwoQubitNorm}, $\rho_\textrm{i}$ denotes the initialized state after initializing laser pulse, $\rho_\textrm{f}$ denotes the final state after applying control sequence to $\rho_\textrm{i}$.
 The RF (RF1) $\pi$ pulse exchanges the population of $|m_S=0, m_I=1\rangle$ and $|m_S=0, m_I=0\rangle$ ($|m_S=0, m_I=0\rangle$ and $|m_S=0, m_I=-1\rangle$).
 The measured $I_{\textrm{PL}}$ after the four sequences ($S_1$, $S_2$, $R_1$, and $R_2$, respectively) satisfy
\begin{equation}
\label{S1-S2}
S_1-S_2 =  (\beta_{|0, 1\rangle}-\beta_{|0, -1\rangle})(P_{|0, 1\rangle}-P_{|0, -1\rangle}),
\end{equation}
\begin{equation}
\label{R1-R2}
R_1-R_2 =  (\beta_{|0, 1\rangle}-\beta_{|0, -1\rangle})(P_{\rho_\textrm{i}, |0, 1\rangle}-P_{\rho_\textrm{i}, |0, -1\rangle}),
\end{equation}
where $P_{|k\rangle}$ ($P_{\rho_\textrm{i}, |k\rangle}$,) is the population of $|k\rangle$ for $\rho_\textrm{f}$ ($\rho_\textrm{i}$). After the initializing laser pulse, the electron spin is polarized with coefficient $\alpha$, and the nuclear spin is almost completely polarized. Thus we have $P_{\rho_\textrm{i}, |0, 1\rangle}=\alpha$ and $P_{\rho_\textrm{i}, |0, -1\rangle}=0$. The state $|m_S=0, m_I=-1\rangle$ remains idle during the control sequence, thus $P_{|0, -1\rangle}=0$. The population of $|m_S=0, m_I=1\rangle$ for the final state can be derived as
\begin{equation}
\label{P01norm}
P_{|01\rangle} = \alpha\frac{S_1-S_2}{R_1-R_2},
\end{equation}

\subsection{Supplementary Note 8}
\textbf{Measuring the polarization of the NV electron spin.} The measurement of the polarization described here is similar to that described before \cite{Nat484_82}. Supplementary Figure \ref{PolarNV} shows the results and pulse sequences used in the measurement.
We first recorded the nuclear Rabi oscillation by driving the $|m_S=0, m_I=1\rangle$ and $|m_S=0, m_I=0\rangle$ transition. The nuclear spin is almost completely polarized. The amplitude of this nuclear Rabi oscillation is proportional to the polarization $\alpha$ of the electron spin, with
\begin{equation}
\label{A1}
A1 =  (\beta_{|0, 1\rangle}-\beta_{|0, 0\rangle})\alpha.
\end{equation}
Secondly, another  nuclear Rabi oscillation is recorded after a MW2 $\pi$ pulse. The MW2 $\pi$ pulse exchanges the population of $|m_S=0, m_I=1\rangle$ and $|m_S=-1, m_I=1\rangle$. The amplitude of this nuclear Rabi oscillation is proportional to the population of $|m_S=-1, m_I=1\rangle$ for the initialized state.
\begin{equation}
\label{A2}
A2 =  (\beta_{|0, 1\rangle}-\beta_{|0, 0\rangle})(1-\alpha)/2,
\end{equation}
Then the polarization $\alpha$ can be obtained with
\begin{equation}
\label{NVpolaralpha}
\alpha =  \frac{1}{1+2\frac{A_2}{A_1}},
\end{equation}
With the results shown in Supplementary Figure \ref{PolarNV}, we estimated the polarization of the NV electron spin to be $\alpha=0.83(2)$.

\subsection{Supplementary Note 9}
\textbf{Measurement of single-qubit average gate fidelity.} We first describe the method for measuring the average gate fidelity of single-qubit gates.
The average gate fidelity of single-qubit gates are measured with randomized benchmarking (RB) method \cite{PRA77_012307}. Unlike that with quantum process tomography, the measured fidelity with RB method is not limited by errors in state preparation and measurement. The qubit is initialized to $|0\rangle$, then a predetermined sequence of randomized computational gates is applied. Each computational gate consists of a Pauli gate followed by a (non-Pauli) Clifford gate. Pauli gates are randomly chosen to rotate the qubit about the $\pm x$, $\pm y$, or $\pm z$ axes for an angle $\pi$ on the Bloch sphere, or to be a $\pm I$ identity gate; Clifford gates are randomly chosen to rotate about the $\pm x$ or $\pm y$ axes for an angle $\pi/2$. The gate sequence is followed by a final Clifford gate chosen to ensure that the final qubit state is $|0\rangle$ if all the gates are ideal. The fidelity of the final state $\rho_\textrm{f}$, $F=\langle0|\rho_\textrm{f}|0\rangle$, is measured. The measured final state fidelity is averaged over different random sequences. The averaged fidelity, $\overline{F}$, is fitted with Eqn. \ref{FRB}
\begin{equation}
\overline{F}= 1/2  + 1/2 (1-d_{\textrm{if}})(1-2\varepsilon_\textrm{g})^{l},
\label{FRB}
\end{equation}
where $l$ is the number of computational gates, $\varepsilon_\textrm{g}$ is the average error per gate, and $d_{\textrm{if}}$ describes errors in state preparation and measurement. The average gate fidelity is
\begin{equation}
F_\textrm{a}= 1-\varepsilon_\textrm{g},
\label{FaveRB}
\end{equation}
In the experiment, $\pm x, \pm y$ rotations are realized by proper microwave settings, and $\pm z$ rotations are implemented by a rotation of the logical frame of the qubit for the subsequent pulses \cite{NJP11_013034, PRA84_030303, PRL113_220501}.

For the naive pulse, each Clifford gate is performed by a rectangular $\pi/2$ pulse and each Pauli gate by a rectangular $\pi$ pulse; For the five-piece SUPCODE pulse, each Clifford gate is performed by a five-piece SUPCODE $2.5\pi$ pulse (equivalent to $\pi/2$ in the single-qubit case, see Section \ref{SingleQubit}) and each Pauli gate by a pair of five-piece SUPCODE $2.5\pi$ pulses; For the BB1 (BB1inC) pulse, each Clifford gate is performed by a BB1 (BB1inC) $\pi/2$ pulse and each Pauli gate by a BB1 (BB1inC) $\pi$ pulse (see Section \ref{SingleQubit}).

The RB results for naive, five-piece SUPCODE,  BB1 and BB1inC pulses are shown in Fig. 2(b) in the main text and summarized in Supplementary Table \ref{RBresults}. The measured average gate fidelities are  0.99968(6), 0.99916(8), 0.999945(6) and 0.999952(6), respectively.

\textrm{\\}

\textbf{Measurement of two-qubit CNOT average gate fidelity.}In the following we describe the method for measuring the average gate fidelity of two-qubit CNOT gate.
The average gate fidelity of CNOT can be measured with Eqn. \ref{FaveCNOT}.
This requires the full knowledge of the quantum operation $\xi$, which is usually very difficult to be obtained.
Quantum process tomography has been developed to characterize the quantum gates.
However, this procedure requires  a number of measurements that scale exponentially with the number of qubits, and the measured process matrix is sensitive to errors in state preparation and measurement.
Randomized bechmarking and related techniques are developed to obtain the average gate fidelities.
However, in the hybrid system composed of electron and nuclear spins, single-qubit gates on the nuclear spin cost longer time than the electron coherence time.
The error of gates on the nuclear spin will dominate the fidelity decay in randomized benchmarking, and the gate fidelity of CNOT can not be precisely determined this way.

Herein, we present a method to estimate the average fidelity of CNOT gate. We determine the fidelity by repeated application of the CNOT gates on the system.
A wealth of information can be obtained by studying the state dynamics under repeated application of quantum gates \cite{NatPhys4_463}.
In Ref. \cite{PRL109_060501}, CNOT gates were repeatedly applied on the input state generated by $X_{-\pi/2}\otimes I$, and the fidelity $F_\textrm{s}$ of final states were measured.
The fidelity $F_\textrm{s}$ decays as the number of the CNOT gate, $N$, is increased. The maximum value of $N$ was $12$ in that work.
By assuming that the decay obeys an exponential model, the gate fidelity $F_\textrm{g}$ can be extracted.

The pulse sequence used in our experiment is shown in the inset of Fig. 4e in the main text.
The initial state of the two-qubit system is prepared by applying a RF $\pi/2$ pulse after the initial laser pulse.
Then $N$, which is even, times of repeated CNOT gates are applied.
Finally, the population of state $|01\rangle$ ($P_{|01\rangle}$) after another RF $\pi /2$ is measured.
Up to $192$ CNOT gates are applied, the dynamics of $P_{|01\rangle}$, however, does not obey a simple exponential decay.
As shown in Supplementary Figure \ref{P01compHamil}, the measured $P_{|01\rangle}$  oscillates while decaying with $N$.
 In our experiment, the nuclear spin qubit is extremely `clean' due to being insensitive to the external noises.
The CNOT gate designed by quantum optimal control method consists of microwave pulses only.
Thus the decay is due to the static fluctuation of $\delta_0$ and $\delta_1$, while the oscillation is mainly due to the deviation of the  experimental operation from the ideal CNOT gate.

We simulated the dynamics of  $P_{|01\rangle}$ based on the Hamiltonian $H_{\textrm{rot}}$, the pulse sequence (shown in Supplementary Figure \ref{GRAPEseqFave}) and the distributions of $\delta_0$ and $\delta_1$. The simulated dynamics of  $P_{|01\rangle}$ is presented as the blue dashed line in Supplementary Figure \ref{P01compHamil}.
The deviation between the experimental result and the simulated result shown in  Supplementary Figure \ref{P01compHamil}, is mainly due to the difference between the $H_{\textrm{rot}}$ and the practical one $H_{\textrm{rot,prac}}$ in the experiment.
The microwave frequency does not equal the resonance frequency exactly, so the off-resonance term $\delta\Omega$ is of a prior unknown nonzero value.
The value of hyperfine coupling strength $A = -2.16~$MHz, which is used for pulse sequence designing, can also deviate slightly from the practical one $A_{\textrm{exp}}$.
We denote this difference as $\delta A = A_{\textrm{exp}}- A$.
The practical Hamiltonian $H_{\textrm{rot,prac}}$  can be extracted by fitting the experimental data.
The fitting procedure is accomplished with Matlab. The best-fit values of the parameters are $\delta A=0.008(1)~$MHz and $\delta\Omega=0.068(4)~$MHz, with the errors being the uncertainty within $95\%$ confidence.
The extracted values of $\delta A$ and $\delta\Omega$ are much smaller than the value of CW spectrum's linewidth.
The fitting result, which agrees with the experimental data well, is shown as the red solid line in Supplementary Figure \ref{P01compHamil}. With the values of  $\delta A$ and $\delta\Omega$ , we can determine the fidelity of CNOT gate according to Eqn. \ref{FaveCNOT}.  The derived average gate fidelity is 0.9920(1), where the error is due to the uncertainty of $\delta A$ and $\delta\Omega$.

\subsection{Supplementary Note 10}
\textbf{Robust and precise optimal control method on NV-NV system.} We have demonstrated a high fidelity CNOT gate at fault-tolerant threshold, taking the NV electron spin and $^{14}$N nuclear spin as qubits. The CNOT gate is designed with modified optimal control method. This method can also be used to design robust and precise quantum gates on NV-NV coupled system, a key ingredient for scalable quantum computation using diamond.

The static Hamiltonian of two coupled NV centers can be described as
\begin{equation}
H_0= H_{\textrm{NV,} 1}+H_{\textrm{NV,} 2}+H_{\textrm{int}},
\label{H0NVNV}
\end{equation}
with
\begin{equation}
H_{\textrm{NV,} 1}= 2\pi DS_{z1}^2-\gamma_e{\bm{B}_{0, 1}}\cdot{\bm{S}_1},
\label{HNV1}
\end{equation}
\begin{equation}
H_{\textrm{NV,} 2}= 2\pi DS_{z2}^2-\gamma_e{\bm{B}_{0, 2}}\cdot{\bm{S}_2},
\label{HNV2}
\end{equation}
\begin{equation}
H_{\textrm{int}}= 2\pi {\bm{S}_1}\cdot{\mathbb{C}}\cdot{\bm{S}_2},
\label{Hint}
\end{equation}
where $\bm{S}_1$ and $\bm{S}_2$ are the spin operators of individual NV centers, NV 1 and NV 2, respectively. The zero filed splitting is $D=$2870~MHz. The coupling tensor between NV 1 and NV 2 is denoted as $\mathbb{C}$. The static magnetic field applied on NV 1 (NV 2) is $\bm{B}_{0, 1}$ ($\bm{B}_{0, 2}$).

The system can be controlled by oscillating magnetic fields. The corresponding control Hamiltonian is
\begin{equation}
\begin{aligned}
&H_{\textrm{C, NV-NV}}(t)= \\
&-\gamma_\textrm{e}\sum_m \cos[2\pi f_mt+\phi_m(t)]\bm{B}_{1, m}(t)\cdot(\bm{S}_1+\bm{S}_2),
\label{HcNVNV}
\end{aligned}
\end{equation}
where $f_m$ are the carrier frequencies of the control fields, $\bm{B}_{1, m}$ contain the amplitudes $B_{1, m}=|\bm{B}_{1, m}|$ and the polarization $\bm{u}_{ m}=\bm{B}_{1, m}/B_{1, m}$. The amplitudes $B_{1, m}$ and the phases $\phi_m$ can be changed in time to steer the system.

We turn to a rotating frame, in which the Hamiltonian is
\begin{equation}
\begin{aligned}
&H'(t)= \\
&e^{i(H_{\textrm{NV,} 1}+H_{\textrm{NV,} 2})t}[H_{\textrm{int}}+H_{\textrm{C, NV-NV}}(t)]e^{-i(H_{\textrm{NV,} 1}+H_{\textrm{NV,} 2})t},
\label{HrotNVNV}
\end{aligned}
\end{equation}
The evolution operator with a time duration $T$ is
\begin{equation}
U_\textrm{e}(T)=\mathcal{T}e^{-i\int_0^TdtH'(t)},
\label{UTNVNV}
\end{equation}
where $\mathcal{T}$ is the time-ordering operator.

Similar to that described in Section \ref{optmcont}, the pulse sequence for a target two-qubit unitary gate $U$ can be designed by maximizing the performance function
\begin{equation}
\label{Fpseq}
F'_{\textrm{seq}} =  \frac{1}{d(d+1)}[tr(M'M'^\dag)+|tr(M')|^2],
\end{equation}
with $d=$4 and
\begin{equation}
\label{Mp}
M' =  U^\dag \mathcal{P}U_\textrm{e}(T)\mathcal{P},
\end{equation}
where $\mathcal{P}$ is projection operator on the two-qubit subspace.

Considering the quasi-static noises from environment and the control fields, the control Hamiltonian in Eqn. \ref{HcNVNV} is replaced by
\begin{equation}
\begin{aligned}
&H_{\textrm{C, err, NV-NV}}(t)= 2\pi\delta_{0, 1}S_{z1}+2\pi\delta_{0, 2}S_{z2} \\
&-\gamma_\textrm{e}\sum_m \cos[2\pi f_mt+\phi_m(t)](1+\delta_{1\textrm{,} m\textrm{, rel}}) \\
&\times\bm{B}_{1, m}(t)\cdot(\bm{S}_1+\bm{S}_2),
\label{HcerrNVNV}
\end{aligned}
\end{equation}
With $H_{\textrm{C, NV-NV}}(t)$ replaced by $H_{\textrm{C, err, NV-NV}}(t)$, it is straight forward to calculate the evolution operator $U_{\textrm{e, err}}(T)$ and gate fidelity $F'_{\textrm{seq, err}}$ with quasi-static noises described by $\delta_{0, 1}$, $\delta_{0, 2}$, and $\delta_{1\textrm{,} m\textrm{, rel}}$. Then the performance function is defined by integrating $F'_{\textrm{seq, err}}$ over distributions of $\delta_{0, 1}$, $\delta_{0, 2}$, and $\delta_{1\textrm{,} m\textrm{, rel}}$. By maximizing the performance function, pulse sequence for target $U$ can be designed to be robust against the noises.

We take the optimization of pulse sequence for a robust and precise CNOT gate as an example. The static magnetic field applied on each NV center is aligned along the NV symmetry axis, and the NV centers can be individually addressable by application of gradient magnetic field. The coupling strength is taken to be $100~$kHz, corresponding to a distance of about 8~nm between two NV centers \cite{NatPhys6_249}. According to Ref. \cite{NatNano9_279}, a magnetic-field gradient of 12~G nm$^{-1}$ is available, corresponding to a difference of more than 200~MHz between the NV centers' resonant frequencies.
 The spin states $|m_S=0\rangle$ and $|m_S=-1\rangle$ of NV 1 (NV 2) are encoded as $|0\rangle$ and $|1\rangle$ of qubit 1 (qubit 2).
 Microwave pulses with two frequencies, which are resonant frequencies for the two qubits, are applied to control the system. The amplitude and phase of each microwave pulses can be modulated to realize the CNOT gate. The CNOT gate is designed so that the state of qubit 2 is flipped iff qubit 1 is in state $|1\rangle$. To make the CNOT gate robust against the noises, we consider the quasi-static distributions of the noises in the optimization of the pulse sequence. The performance function is defined as the average gate fidelity
\begin{equation}
\begin{aligned}
&F'_\textrm{a}=\int d\delta_{0, 1}\int d\delta_{1\textrm{,} 1\textrm{, rel}}\int d\delta_{0, 2}\int d\delta_{1\textrm{,} 2\textrm{, rel}} \\
& \times P_{0, 1}(\delta_{0, 1})P_{1\textrm{,} 1\textrm{, rel}}(\delta_{1\textrm{,} 1\textrm{, rel}})P_{0, 2}(\delta_{0, 2})P_{1\textrm{,} 2\textrm{, rel}}(\delta_{1\textrm{,} 2\textrm{, rel}}) \\
& \times F'_{\textrm{seq, err}},
\label{Fpa}
\end{aligned}
\end{equation}
where $P_{0, 1}(\delta_{0, 1})$ and $P_{1\textrm{,} 1\textrm{, rel}}(\delta_{1\textrm{,} 1\textrm{, rel}})$ ($P_{0, 2}(\delta_{0, 2})$ and $P_{1\textrm{,} 2\textrm{, rel}}(\delta_{1\textrm{,} 2\textrm{, rel}})$) describe the distributions of quasi-static dephasing noise and control field fluctuation on NV 1 (NV 2).
The distribution $P_{0, 1}(\delta_{0, 1})$ ($P_{0, 2}(\delta_{0, 2})$) is mainly due to  the $^{13}$C nuclear spin bath surrounding the NV 1 (NV 2).
 Since $^{13}$C is naturally abundant,  the distribution $P_0(\delta_{0})$ in the main text is a typical estimation of $P_{0, 1}(\delta_{0, 1})$ ($P_{0, 2}(\delta_{0, 2})$).
 The distributions $P_{1\textrm{,} 1\textrm{, rel}}(\delta_{1\textrm{,} 1\textrm{, rel}})$ and $P_{1\textrm{,} 2\textrm{, rel}}(\delta_{1\textrm{,} 2\textrm{, rel}})$, mainly depending on the microwave generator, are consistent with $P_1(\delta_1)$ which can be obtained from experiment data. Considering the distributions, a pulse sequence for the CNOT gate can be optimized to achieve an average gate fidelity $F'_\textrm{a}=0.9926$ by our method. The pulse sequence is shown in Supplementary Figure \ref{NVNVCNOTseq}. Thus our method can be applied to realize \emph{robust} and \emph{high fidelity }two-qubit gate on spatially separated NV centers.

\end{document}